\documentstyle[graphics]{cupconf}  
  
\ifoldfss  
\else  
  \ifnfssone  
    \newmathalphabet{\mathit}  
      \addtoversion{normal}{\mathit}{cmr}{m}{it}  
      \addtoversion{bold}{\mathit}{cmr}{bx}{it}  
    \newmathalphabet{\mathcal}  
      \addtoversion{normal}{\mathcal}{cmsy}{m}{n}  
    \else  
    \ifnfsstwo  
    \fi  
  \fi  
\fi  
  
\def\gtrsim{\mathrel{\hbox{\rlap{\hbox{\lower4pt\hbox{$\sim$}}}\hbox{$>$}}}}  
\def\lesssim{\mathrel{\hbox{\rlap{\hbox{\lower4pt\hbox{$\sim$}}}\hbox{$<$}}}}

\def\hexnumber#1{\ifcase#1 0\or1\or2\or3\or4\or5\or6\or7\or8\or9\or  
 A\or B\or C\or D\or E\or F\fi }  
  
\makeatletter  
\ifx\CUP@mtlplain@loaded\undefined  
\else  
\fi  
\makeatother  
 \makeatletter  
 \ifx\CUP@mtlplain@loaded\undefined  
   \font\tenbmi=cmmib10 at 10pt  
   \font\sevenbmi=cmmib10 at 7pt  
   \font\fivebmi=cmmib10 at 5pt  
  
   \newfam\bmifam  
   \textfont\bmifam=\tenbmi  
   \scriptfont\bmifam=\sevenbmi  
   \scriptscriptfont\bmifam=\fivebmi  
     
 \fi  
 \makeatother  
\ifnfsstwo

\fi  
\ifnfssone

\fi  
\ifoldfss

\fi

\mathchardef\varLambda="0103

\makeatletter  
\ifx\CUP@mtlplain@loaded\undefined  
\else  
\fi  
\makeatother  
\makeatletter  
\ifx\CUP@mtlplain@loaded\undefined  
  \font\tenbms=cmbsy10  
  \font\sevenbms=cmbsy10 at 7pt  
  \font\fivebms=cmbsy10 at 5pt  
  \newfam\bmsfam  
  \textfont\bmsfam=\tenbms  
  \scriptfont\bmsfam=\sevenbms  
  \scriptscriptfont\bmsfam=\fivebms  

  \edef\bsy@{\hexnumber\bmsfam}  
  \mathchardef\bnabla="0\bsy@72  
\fi  
\makeatother  
\def\eg{{e.g.\ }}  
  
\def\etal{\mbox{\it et al.}}

\newcommand{\be}{\begin{equation}} 
\newcommand{\ee}{\end{equation}} 
\newcommand{\Deln}{$\Delta N_\nu\;$} 
\newcommand{\epm}{$e^{\pm}\;$} 
 
\def\eg{{\it e.g.},~} 
\def\4he{$^4$He} 
\def\3he{$^3$He} 
\def\7li{$^7$Li} 
\def\Yp{Y$_{\rm P}$~} 
\def\hii{H\thinspace{$\scriptstyle{\rm II}$}~} 
\def\h1{H\thinspace{$\scriptstyle{\rm I}$}} 
\def\d1{D\thinspace{$\scriptstyle{\rm I}$}} 
\def\etal{{\it et al.}~} 
\newcommand\la{\lower0.6ex\vbox{\hbox{$ \buildrel{\textstyle  
<}\over{\sim}\ $}}} 
\newcommand\ga{\lower0.6ex\vbox{\hbox{$ \buildrel{\textstyle  
>}\over{\sim}\ $}}} 
\newcommand{\obh}{$\Omega_{\rm B} h^2\;$} 
 
\title[Primordial Alchemy]{Primordial Alchemy:\\  
From The Big Bang To The Present Universe}  
  
\author[Gary Steigman]%
{G\ls A\ls R\ls Y\ns S\ls T\ls E\ls I\ls G\ls M\ls A\ls N}  
  
\affiliation{Departments of Physics and Astronomy,\\ 
The Ohio State University, Columbus, OH 43210, USA\\  
}  
  
\begin{document}  
\ifnfssone  
\else  
  \ifnfsstwo  
  \else  
    \ifoldfss  
      \let\mathcal\cal  
      \let\mathrm\rm  
      \let\mathsf\sf  
    \fi  
  \fi  
\fi  
  
\maketitle  
  
  
\begin{abstract}  
 
Of the light nuclides observed in the universe today, D, 
\3he, \4he, and \7li are relics from its early evolution. 
The primordial abundances of these relics, produced via Big Bang  
Nucleosynthesis (BBN) during the first half hour of the evolution  
of the universe provide a unique window on Physics and Cosmology  
at redshifts $\sim 10^{10}$.  Comparing the BBN-predicted abundances 
with those inferred from observational data tests the consistency 
of the standard cosmological model over ten orders of magnitude in 
redshift, constrains the baryon and other particle content of the 
universe, and probes both Physics and Cosmology beyond the current 
standard models.  These lectures are intended to introduce students, 
both of theory and observation, to those aspects of the evolution of  
the universe relevant to the production and evolution of the light  
nuclides from the Big Bang to the present.  The current observational  
data is reviewed and compared with the BBN predictions and the 
implications for cosmology (\eg universal baryon density) and  
particle physics (\eg relativistic energy density) are discussed. 
While this comparison reveals the stunning success of the standard 
model(s), there are currently some challenges which leave open the 
door for more theoretical and observational work with potential 
implications for astronomy, cosmology, and particle physics. 
 
\end{abstract}  
 
The present universe is expanding and is filled with radiation 
(the 2.7~K Cosmic Microwave Background -- CMB) as well as 
``ordinary" matter (baryons), ``dark" matter and, ``dark energy". 
Extrapolating back to the past, the early universe was hot and 
dense, with the overall energy density dominated by relativistic 
particles (``radiation dominated").  During its early evolution 
the universe hurtled through an all too brief epoch when it served 
as a primordial nuclear reactor, leading to the synthesis of the 
lightest nuclides: D, \3he, \4he, and \7li.  These relics from 
the distant past provide a unique window on the early universe, 
probing our standard models of cosmology and particle physics. 
By comparing the predicted primordial abundances with those 
inferred from observational data we may test the standard 
models and, perhaps, uncover clues to modifications or 
extensions beyond them. 
  
These notes summarize the lectures delivered at the XIII Canary  
Islands Winter School of Astrophysics: ``Cosmochemistry: The Melting  
Pot of Elements''.  The goal of the lectures was to provide both  
theorists and observers with an overview of the evolution of the  
universe from its earliest epochs to the present, concentrating on 
the production, evolution, and observations of the light nuclides. 
Standard Big Bang Nucleosynthesis (SBBN) depends on only one free 
parameter, the universal density of baryons; fixing the primordial 
abundances fixes the baryon density at the time of BBN.  But, since 
baryons are conserved (at least for these epochs), fixing the baryon 
density at a redshift $\sim 10^{10}$, fixes the present-universe 
baryon density.  Comparing this prediction with other, independent 
probes of the baryon density in the present and recent universe 
offers the opportunity to test the consistency of our standard, 
hot big bang cosmological model.   
 
Since these lectures are intended for a student audience, they  
begin with an overview of the physics of the early evolution of  
the Universe in the form of a ``quick and dirty" mini-course on 
Cosmology (\S1).  The experts may wish to skip this material.   
With the necessary background in place, the second lecture (\S2)  
discusses the physics of primordial nucleosynthesis and outlines  
the abundances predicted by the standard model.  The third lecture  
considers the evolution of the abundances of the relic nuclides  
from BBN to the present and reviews the observational status of  
the primordial abundances (\S3).  As is to be expected in such  
a vibrant and active field of research, this latter is a moving  
target; the results presented here represent the status in November  
2001.  Armed with the predictions and the observations, the fourth  
lecture (\S4) is devoted to the confrontation between them.  As  
is by now well known, this confrontation is a stunning success  
for SBBN.  However, given the precision of the predictions and  
of the observational data, it is inappropriate to ignore some  
of the potential discrepancies.  In the end, these may be traceable 
to overly optimistic error budgets, to unidentified systematic  
errors in the abundance determinations, to incomplete knowledge  
of the evolution from the big bang to the present or, to new  
physics beyond the standard models.  In the last lecture I present  
a selected overview of BBN in some non-standard models of Cosmology  
and Particle Physics (\S5).  Although I have attempted to provide  
a {\it representative} set of references, I am aware they are  
incomplete and I apologize in advance for any omissions.

\section{The Early Evolution of the Universe}  
 
Observations of the present universe establish that, on sufficiently 
large scales, galaxies and clusters of galaxies are distributed 
homogeneously and they are expanding isotropically.  On the assumption  
that this is true for the large scale universe throughout its evolution  
(at least back to redshifts $\sim 10^{10}$, when the universe was a  
few hundred milliseconds old), the relation between space-time points  
may be described uniquely by the Robertson -- Walker Metric 
\begin{equation} 
ds^{2} = c^{2}dt^{2} - a^{2}(t)({dr^{2} \over 1 - {\kappa}r^{2}} +  
r^{2}d\Omega^{2})\,, 
\label{rwmetric} 
\end{equation} 
where $r$ is a {\it comoving} radial coordinate and $\theta$ 
and $\phi$ are {\it comoving} spherical coordinates related by 
\begin{equation} 
d{\Omega}^{2} \equiv d{\theta}^{2} + sin^{2}{\theta}d{\phi}^{2}\,. 
\end{equation} 
A useful alternative to the comoving radial coordinate 
$r$ is $\Theta$, defined by 
\be 
d\Theta \equiv {dr \over (1 - {\kappa}r^{2})^{1/2}}\,. 
\ee 
The 3-space curvature is described by $\kappa$, the curvature constant. 
For closed (bounded), or ``spherical" universes, $\kappa > 0$; for  
open (unbounded), or ``hyperbolic" models, $\kappa < 0$; when $\kappa  
= 0$, the universe is spatially flat or ``Euclidean".  It is the  
``scale factor", $a = a(t)$, which describes how physical distances  
between comoving locations change with time.  As the universe expands,  
$a$ increases while, for comoving observers, $r$, $\theta$, and $\phi$ 
remain fixed.  The growth of the separation between comoving observers 
is solely due to the growth of $a$.  Note that {\bf neither} $a$ {\bf  
nor} $\kappa$ is observable since a rescaling of $\kappa$ can always  
be compensated by a rescaling of $a$. 
 
Photons and other massless particles travel on geodesics: $ds = 0$;  
for them (see eq.~\ref{rwmetric}) $d\Theta = \pm ~cdt/a(t)$.  To  
illustrate the significance of this result consider a photon travelling  
from emission at time $t_{e}$ to observation at a later time $t_{o}$.   
In the course of its journey through the universe the photon traverses  
a comoving radial distance $\Delta\Theta$, where 
\be 
\Delta\Theta = \int^{t_{o}}_{t_{e}} ~{cdt \over a(t)}\,. 
\ee 
 
Some special choices of $t_{e}$ or $t_{o}$ are of particular interest. 
For $t_{e} \rightarrow 0$, $\Delta\Theta \equiv \Theta_{\rm H}(t_{o})$  
is the comoving radial distance to the ``Particle Horizon" at time $t_0$.  
It is the comoving distance a photon could have travelled (in the absence 
of scattering or absorption) from the beginning of the expansion of the 
universe until the time $t_{o}$.  The ``Event Horizon", $\Theta_{\rm E}
(t_{e})$, corresponds to the limit $t_{o} \rightarrow \infty$ (provided  
that $\Theta_{\rm E}$ is finite!).  It is the comoving radial distance  
a photon will travel for the entire future evolution of the universe, 
after it is emitted at time $t_{e}$.   
 
\subsection{Redshift} 
 
Light emitted from a comoving galaxy located at $\Theta_{g}$ at time  
$t_{e}$ will reach an observer situated at $\Theta_{o} \equiv 0$ at 
a later time $t_{o}$, where 
\be 
\Theta_{g}(t_{o}, t_{e}) = \int^{t_{o}}_{t_{e}} ~{cdt \over a(t)}\,. 
\label{thetag} 
\ee 
Equation \ref{thetag} provides the relation among $\Theta_{g}$, $t_{o}$,  
and $t_{e}$.  For a comoving galaxy, $\Theta_{g}$ is unchanged so that 
differentiating eq.~\ref{thetag} leads to 
\be 
{dt_{o} \over dt_{e}} = {a_{o} \over a_{e}} = {\nu_{e} \over \nu_{o}}  
= {\lambda_{o} \over \lambda_{e}}\,. 
\ee 
This result relates the evolution of the universe ($a_{o}/a_{e}$) as 
the photon travels from emission to observation, to the change in its 
frequency ($\nu$) or wavelength ($\lambda$).  As the universe expands 
(or contracts!), wavelengths expand (contract) and frequencies decrease 
(increase).  The redshift of a spectral line is defined by relating 
the wavelength at emission (the ``lab" or ``rest-frame" wavelength  
$\lambda_{e}$) to the wavelength observed at a later time $t_{o}$,  
$\lambda_{o}$. 
\be 
z \equiv {\lambda_{o} - \lambda_{e} \over \lambda_{e}} ~\Longrightarrow  
~1 + z = {a_{o} \over a_{e}} = {\nu_{e} \over \nu_{o}}\,. 
\label{redshift} 
\ee 
 
Since the energies of photons are directly proportional to their 
frequencies, as the universe expands photon energies redshift to 
smaller values: E$_{\gamma} = h\nu \Longrightarrow ~$E$_{\gamma}  
\propto (1 + z)^{-1}$.  For {\bf all} particles, massless or not, 
de Broglie told us that wavelength and momentum are inversely related, 
so that: p ~$\propto ~\lambda^{-1} \Longrightarrow ~$p$ ~\propto  
~(1 + z)^{-1}$.  All momenta redshift; for non-relativistic  
particles (\eg galaxies) this implies that their ``peculiar"  
velocities redshift: v = p/M $\propto (1 + z)^{-1}$. 
 
\subsection{Dynamics} 
 
Everything discussed so far has been ``geometrical", relying only on  
the form of the Robertson-Walker metric.  To make further progress  
in understanding the evolution of the universe, it is necessary to  
determine the time dependence of the scale factor $a(t)$.  Although 
the scale factor is not an observable, the expansion rate, the Hubble  
parameter, $H = H(t)$, is. 
\be 
H(t) \equiv {1 \over a}({da \over dt})\,. 
\ee 
The present value of the Hubble parameter, often referred to as the 
Hubble ``constant", is $H_{0} \equiv H(t_{0}) \equiv 100~h$~km~s$^
{-1}$Mpc$^{-1}$ (throughout, unless explicitly stated otherwise, 
the subscript ``0" indicates the present time).  The inverse of the  
Hubble parameter provides an expansion timescale, $H_{0}^{-1} =  
9.78~h^{-1}$~Gyr.  For the HST Key Project (Freedman \etal 2001)  
value of $H_{0} = 72$~km~s$^{-1}$Mpc$^{-1}$ ($h = 0.72$),  
$H_{0}^{-1} = 13.6$~Gyr.   
 
The time-evolution of $H$ describes the evolution of the universe. 
Employing the Robertson-Walker metric in the Einstein equations of 
General Relativity (relating matter/energy content to geometry) leads 
to the Friedmann equation 
\be 
H^{2} = {8\pi \over 3}G\rho - {\kappa c^{2} \over a^{2}}\,. 
\label{friedmann} 
\ee 
It is convenient to introduce a {\it dimensionless} density parameter, 
$\Omega$, defined by  
\be  
\Omega \equiv {8\pi G\rho \over 3H^{2}}\,. 
\label{omega} 
\ee 
We may rearrange eq.~\ref{friedmann} to highlight the relation between 
matter content and geometry 
\be 
\kappa c^{2} = (aH)^{2}(\Omega - 1)\,. 
\label{kappa} 
\ee 
Although, in general, $a$, $H$, and $\Omega$ are all time-dependent, 
eq.~\ref{kappa} reveals that if ever $\Omega < 1$, then it will always 
be $< 1$ {\bf and} in this case the universe is open ($\kappa < 0$). 
Similarly, if ever $\Omega > 1$, then it will always be $> 1$ {\bf and}  
in this case the universe is closed ($\kappa > 0$).  For the special 
case of $\Omega = 1$, where the density is equal to the ``critical 
density" $\rho_{\rm crit} \equiv 3H^{2}/8\pi G$, $\Omega$ is always 
unity and the universe is flat (Euclidean 3-space sections; $\kappa  
= 0$). 
 
The Friedmann equation (eq.~\ref{friedmann}) relates the time-dependence 
of the scale factor to that of the density.  The Einstein equations  
yield a second relation among these which may be thought of as the 
surrogate for energy conservation in an expanding universe. 
\be 
{d\rho \over \rho} + 3(1 + {p \over \rho}){da \over a} = 0\,. 
\label{eqnofstate} 
\ee 
For ``matter" (non-relativistic matter; often called ``dust"), 
$p \ll \rho$, so that $\rho/\rho_{0} = (a_{0}/a)^{3}$.  In contrast,  
for ``radiation" (relativistic 
particles) $p = \rho/3$, so that $\rho/\rho_{0} = (a_{0}/a)^{4}$. 
Another interesting case is that of the energy density and pressure 
associated with the vacuum (the quantum mechanical vacuum is not 
empty!).  In this case $p = -\rho$, so that $\rho = \rho_{0}$. 
This provides a term in the Friedmann equation entirely equivalent 
to Einstein's ``cosmological constant" $\Lambda$.  More generally,  
for $p = w\rho$, $\rho/\rho_{0} = (a_{0}/a)^{3(1+w)}$. 
 
Allowing for these three contributions to the total energy density, 
eq.~\ref{friedmann} may be rewritten in a convenient dimensionless 
form 
\be 
({H \over H_{0}})^{2} = \Omega_{\rm M}({a_{0} \over a})^{3} +  
\Omega_{\rm R}({a_{0} \over a})^{4} + \Omega_{\rm \Lambda} +  
(1 - \Omega)({a_{0} \over a})^{2}\,, 
\label{hsquared} 
\ee 
where $\Omega \equiv \Omega_{\rm M} + \Omega_{\rm R} +  
\Omega_{\rm \Lambda}$. 
 
Since our universe is expanding, for the early universe ($t \ll  
t_{0}$) $ a \ll a_{0}$, so that it is the ``radiation" term in 
eq.~\ref{hsquared} which dominates; the early universe is 
radiation-dominated (RD).  In this case $a \propto t^{1/2}$  
and $\rho \propto t^{-2}$, so that the age of the universe or,  
equivalently, its expansion rate is fixed by the radiation  
density.  For thermal radiation, the energy density is only 
a function of the temperature ($\rho_{\rm R} \propto T^{4}$). 
 
\subsubsection{Counting Relativistic Degrees of Freedom} 
 
It is convenient to write the total (radiation) energy density 
in terms of that in the CMB photons 
\be 
\rho_{\rm R} \equiv ({g_{eff} \over 2})\rho_{\gamma}\,, 
\ee 
where $g_{eff}$ counts the ``effective" relativistic degrees 
of freedom.  Once $g_{eff}$ is known or specified, the  
time -- temperature relation is determined.  If the temperature 
is measured in energy units ($kT$), then 
\be 
t({\rm sec}) = ({2.4 \over g_{eff}^{1/2}})T^{-2}_{\rm MeV}\,. 
\ee 
If more relativistic particles are present, $g_{eff}$ increases 
and the universe would expand faster so that, at {\bf fixed} $T$,  
the universe would be younger.  Since the synthesis of the elements  
in the expanding universe involves a competition between reaction 
rates and the universal expansion rate, $g_{eff}$ will play a key  
role in determining the BBN-predicted primordial abundances. 
 
\begin{itemize} 
 
\item{\it Photons}  
 
Photons are vector bosons.  Since they are massless, they have  
only two degress of freedom: $g_{eff} = 2$.  At temperature $T$  
their number density is $n_{\gamma} = 411(T/2.726K)^{3}~$cm$^{-3}  
= 10^{31.5}T^{3}_{\rm MeV}~$cm$^{-3}$, while their contribution  
to the total radiation energy density is $\rho_{\gamma} = 0.261 
(T/2.726K)^{4}$~eV~cm$^{-3}$.  Taking the ratio of the energy  
density to the number density leads to the average energy per  
photon $\langle {\rm E}_{\gamma} \rangle = \rho_{\gamma}/n_{\gamma}  
= 2.70~kT$.  All other relativistic {\bf bosons} may be simply  
related to photons by 
\be 
{n_{\rm B} \over n_{\gamma}} = {g_{\rm B} \over 2}({T_{\rm B}  
\over T_{\gamma}})^{3} \,, ~~~~~~{\rho_{\rm B} \over \rho_{\gamma}}  
= {g_{\rm B} \over 2}({T_{\rm B} \over T_{\gamma}})^{4}\,,  
~~~~~~\langle {\rm E}_{\rm B} \rangle = 2.70~kT_{\rm B}\,. 
\ee 
The $g_{\rm B}$ are the boson degrees of freedom (1 for a scalar,  
2 for a vector, etc.).  In general, some bosons may have decoupled  
from the radiation background and, therefore, they will not  
necessarily have the same temperature as do the photons ($T_{\rm B}  
\neq T_{\gamma}$). 
 
\item{Relativistic Fermions} 
 
Accounting for the difference between the Fermi-Dirac and Bose-Einstein 
distributions, relativistic fermions may also be related to photons 
\be 
{n_{\rm F} \over n_{\gamma}} = {3 \over 4}{g_{\rm F} \over 2}({T_{\rm F}  
\over T_{\gamma}})^{3} \,, ~~~~~~{\rho_{\rm F} \over \rho_{\gamma}}  
= {7 \over 8}{g_{\rm F} \over 2}({T_{\rm F} \over T_{\gamma}})^{4}\,,  
~~~~~~\langle {\rm E}_{\rm F} \rangle = 3.15~kT_{\rm F}\,. 
\ee 
$g_{\rm F}$ counts the fermion degrees of freedom.  For example, for 
electrons (spin up, spin down, electron, positron) $g_{\rm F} = 4$, 
while for neutrinos (lefthanded neutrino, righthanded antineutrino) 
$g_{\rm F} = 2$. 
 
\end{itemize} 
 
Accounting for all of the particles present at a given epoch in the 
early (RD) evolution of the universe, 
\be 
g_{eff} ~= ~\Sigma_{\rm B}~g_{\rm B}({T_{\rm B} \over T_{\gamma}})^{4}  
~+ ~{7 \over 8}~\Sigma_{\rm F}~g_{\rm F}({T_{\rm F} \over  
T_{\gamma}})^{4}\,. 
\ee 
For example, for the standard model particles at temperatures 
$T_{\gamma} \approx $~few MeV there are photons, electron-positron 
pairs, and three ``flavors" of lefthanded neutrinos (along with 
their righthanded antiparticles).  At this stage all these particles 
are in equilibrium so that $T_{\gamma} = T_{e} = T_{\nu}$ where 
$\nu \equiv \nu_{e}$, $\nu_{\mu}$, $\nu_{\tau}$.  As a result 
\be 
g_{eff} = 2 + {7 \over 8}(4 + 3 \times 2) = {43 \over 4}\,, 
\ee 
leading to a time -- temperature relation: $t = 0.74~T^{-2} 
_{\rm Mev}$~sec. 
 
As the universe expands and cools below the electron rest mass energy, 
the \epm pairs annihilate, heating the CMB photons, but {\bf not} the 
neutrinos which have already decoupled.  The decoupled neutrinos 
continue to cool by the expansion of the universe ($T_{\nu} \propto  
a^{-1}$), as do the photons which now have a higher temperature  
$T_{\gamma} = (11/4)^{1/3}T_{\nu}$ ($n_{\gamma}/n_{\nu} = 11/3$). 
During these 
epochs 
\be 
g_{eff} = 2 + {7 \over 8} \times 3 \times 2({4 \over 11})^{4/3}  
= 3.36\,, 
\ee 
leading to a modified time -- temperature relation: $t =  
1.3~T^{-2}_{\rm Mev}$~sec. 
 
\subsubsection{``Extra" Relativistic Energy} 
 
Suppose there is some new physics beyond the standard model of 
particle physics which leads to ``extra" relativistic energy so 
that $\rho_{\rm R} \rightarrow \rho_{\rm R}' \equiv \rho_{\rm R}  
+ \rho_{X}$; hereafter, for convenience of notation, the subscript 
R will be dropped.  It is useful, and conventional, to account 
for this extra energy in terms of the equivalent number of extra 
neutrinos: $\Delta N_{\nu} \equiv \rho_{X}/\rho_{\nu}$ (Steigman, 
Schramm, \& Gunn 1977 (SSG); see also Hoyle \& Tayler 1964, Peebles 
1966, Shvartsman 1969).  In the presence of this extra energy, 
prior to \epm annihilation 
\be 
{\rho' \over \rho_{\gamma}} = {43 \over 8}~(1 + {7\Delta N_{\nu}  
\over 43}) = 5.375~(1 + 0.1628~\Delta N_{\nu})\,. 
\ee 
In this case the early universe would expand faster than in  
the standard model.  The pre-\epm annihilation speedup in the 
expansion rate is 
\be 
S_{pre} \equiv {t \over t'} = ({\rho' \over \rho})^{1/2} =  
(1 + 0.1628~\Delta N_{\nu})^{1/2}\,. 
\ee 
 
After \epm annihilation there are similar, but quantitatively 
different changes 
\be 
{\rho' \over \rho_{\gamma}} = 1.681~(1 + 0.1351~\Delta N_{\nu})\,,  
~~~~~~~~S_{post} = (1 + 0.1351~\Delta N_{\nu})^{1/2}\,. 
\ee 
 
Armed with an understanding of the evolution of the early universe 
and its particle content, we may now proceed to the main subject of 
these lectures, primordial nucleosynthesis. 
 
\section{Big Bang Nucleosynthesis and the Primordial Abundances} 
 
Since the early universe is hot and dense, interactions among the 
various particles present are rapid and equilibrium among them is  
established quickly.  But, as the universe expands and cools, there 
are departures from equilibrium; these are at the core of the most 
interesting themes of our story.   
 
\subsection{An Early Universe Chronology} 
 
At temperatures above a few MeV, when the universe is tens of 
milliseconds old, interactions among photons, neutrinos, electrons,  
and positrons establish and maintain equilibrium ($T_{\gamma} =  
T_{\nu} = T_{e}$).  When the temperature drops below a few MeV the  
weakly interacting neutrinos decouple, continuing to cool and dilute  
along with the expansion of the universe ($T_{\nu} \propto a^{-1}$,  
$n_{\nu} \propto T_{\nu}^{3}$, and $\rho_{\nu} \propto T_{\nu}^{4}$). 
 
\subsubsection{Neutron -- Proton Interconversion} 
 
Up to now we haven't considered the baryon (nucleon) content of 
the universe.  At these early times there are neutrons and protons 
present whose relative abundance is determined by the usual weak  
interactions. 
\be 
p + e^{-} ~\Longleftrightarrow ~n + \nu_{e}\,, ~~~~n + e^{+} 
~\Longleftrightarrow ~p + \bar{\nu}_{e}\,, ~~~~n ~\Longleftrightarrow  
~p + e^{-} + \bar{\nu}_{e}\,. 
\label{betadecay} 
\ee 
As time goes by and the universe cools, the lighter protons are favored 
over the heavier neutrons and the neutron-to-proton ratio decreases, 
initially as $n/p \propto $~exp$(-\Delta m/T)$, where $\Delta m = 1.29$  
MeV is the neutron-proton mass difference.  As the temperature drops 
below roughly 0.8 MeV, when the universe is roughly one second old, 
the rate of the two-body collisions in eq.~\ref{betadecay} becomes 
slow compared to the universal expansion rate and deviations from 
equilibrium occur.  This is often referred to as ``freeze-out", but 
it should be noted that the $n/p$ ratio continues to decrease as the 
universe expands, albeit at a slower rate than if the ratio tracked 
the exponential.  Later, when the universe is several hundred seconds 
old, a time comparable to the neutron lifetime ($\tau_{n} = 885.7 \pm  
0.8$~sec.), the $n/p$ ratio resumes falling exponentially: $n/p \propto  
$~exp$(-t/\tau_{n})$.  Notice that the $n/p$ ratio at BBN depends on 
the competition between the weak interaction rates and the early 
universe expansion rate so that any deviations from the standard 
model (\eg $\rho \rightarrow \rho + \rho_{X}$) will change the 
relative numbers of neutrons and protons available for building 
more complex nuclides.  
 
\subsubsection{Building The Elements} 
 
At the same time that neutrons and protons are interconverting,  
they are also colliding among themselves to create deuterons:  
$n + p \Longleftrightarrow D + \gamma$.  However, at early  
times when the density and average energy of the CMB photons is  
very high, the newly-formed deuterons find themselves bathed in  
a background of high energy gamma rays capable of photodissociating 
them.  As we shall soon see, there are more than a billion photons 
for every nucleon in the universe so that before a neutron or a 
proton can be added to D to begin building the heavier nuclides, 
the D is photodissociated.  This bottleneck to BBN beginning in 
earnest persists until the temperature drops sufficiently so that 
there are too few photons energetic enough to photodissociate the 
deuterons before they can capture nucleons to launch BBN.  This 
occurs after \epm annihilation, when the universe is a few minutes 
old and the temperature has dropped below 80 keV (0.08 MeV). 
 
Once BBN begins in earnest, neutrons and protons quickly combine to 
form D, $^3$H, \3he, and \4he.  Here, there is another, different 
kind of bottleneck.  There is a gap at mass-5; there is no 
stable mass-5 nuclide.  To jump the gap requires \4he reactions 
with D or $^3$H or \3he, all of which are positively charged. 
The coulomb repulsion among these colliding nuclei suppresses 
the reaction rate ensuring that virtually all of the neutrons 
available for BBN are incorporated in \4he (the most tightly  
bound of the light nuclides), and also that the abundances 
of the heavier nuclides are severely depressed below that of \4he 
(and even of D and \3he).  Recall that $^3$H is unstable and will 
decay to \3he.  The few reactions which manage to bridge the mass-5 
gap mainly lead to mass-7 (\7li, or $^7$Be which later, when the 
universe has cooled further, will capture an electron and decay 
to \7li); the abundance of $^6$Li is below that of the more tightly 
bound \7li by one to two orders of magnitude.  There is another 
gap at mass-8.  This absence of any stable mass-8 nuclides  
ensures there will be no astrophysically interesting production 
of heavier nuclides. 
 
The primordial nuclear reactor is short-lived, quickly encountering 
an energy crisis.  Because of the falling temperatures and the 
coulomb barriers, nuclear reactions cease rather abruptly when 
the temperature drops below roughly 30 keV, when the universe is 
about 20 minutes old.  As a result there is ``nuclear freeze-out" 
since no already existing nuclides are destroyed (except for those 
that are unstable and decay) and no new nuclides are created.  In 
$\sim 1000$ seconds BBN has run its course. 
 
\subsection{The SBBN-Predicted Abundances} 
 
The primordial abundances of D, \3he, and \7li($^7$Be) are rate 
limited, depending sensitively on the competition between the 
nuclear reactions rates and the universal expansion rate.  As 
a result, these nuclides are potential baryometers since their 
abundances are sensitive to the universal density of nucleons. 
As the universe expands, the nucleon density decreases so it 
is useful to compare the nucleon density to that of the CMB 
photons $\eta \equiv n_{\rm N}/n_{\gamma}$.  Since this ratio 
will turn out to be very small, it is convenient to introduce 
\be 
\eta_{10} \equiv 10^{10}(n_{\rm N}/n_{\gamma}) = 274\Omega_ 
{\rm B}h^{2}\,. 
\ee 
As the universe evolves (post-\epm annihilation) this ratio 
is accurately preserved so that $\eta_{\rm BBN} = \eta_{0}$. 
Testing this relation over ten orders of magnitude in redshift, 
over a range of some ten billion years, can provide a confirmation 
of or a challenge to the standard model.  
 
In contrast to the other light nuclides, the primordial abundance 
of \4he (mass fraction Y) is relatively insensitive to the baryon 
density, but since virtually all neutrons available at BBN are  
incorporated in \4he, it does depend on the competition between  
the weak interaction rate (largely fixed by the accurately measured  
neutron lifetime) and the universal expansion rate (which depends 
on $g_{eff}$).  The higher the nucleon density, the earlier can  
the D-bottleneck be breached.  At early times there are more 
neutrons and, therefore, more \4he will be synthesized.  This 
latter effect is responsible for the very slow (logarithmic) 
increase in Y with $\eta$.  Given the standard model relation 
between time and temperature and the nuclear and weak cross 
sections and decay rates measured in the laboratory, the evolution 
of the light nuclide abundances may be calculated and the frozen-out 
relic abundances predicted as a function of the one free parameter,  
the nucleon density or $\eta$.  These are shown in Figure 1. 
 
\begin{figure}[t!]\label{schrammplot}  
\resizebox{\hsize}{!}{\includegraphics{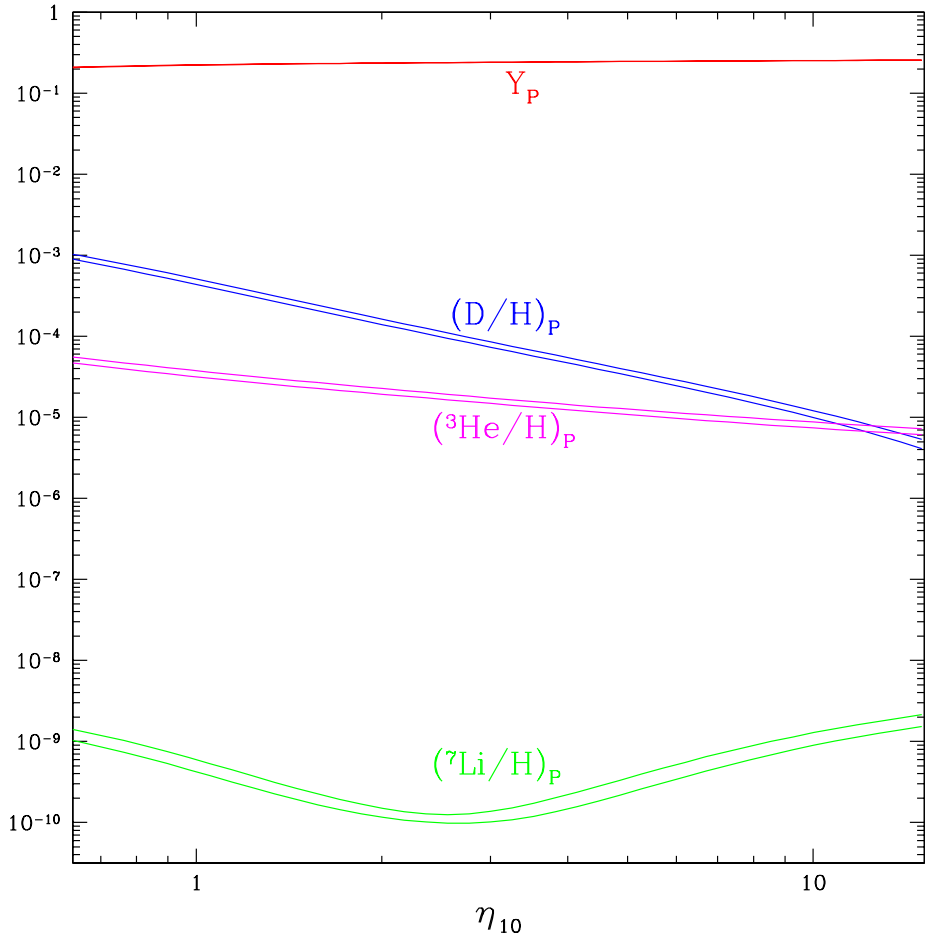}}  
\hfill  
\parbox[b]{\hsize}{  
\caption{The SBBN-predicted primordial abundances of D, \3he, 
\7li (by number with respect to hydrogen), and the \4he mass  
fraction Y as a function of the nucleon abundance $\eta_{10}$.   
The widths of the bands reflect the theoretical uncertainties.}}  
\end{figure}  
 
Not shown on Figure 1 are the relic abundances of $^6$Li, $^9$Be, 
$^{10}$B, and $^{11}$B, all of which, over the same range in $\eta$, 
lie offscale, in the range $10^{-20} - 10^{-13}$.   
 
The reader may notice the abundances appear in Figure 1 as bands.   
These represent the theoretical uncertainties in the predicted  
abundances.  For D/H and \3he/H they are at the $\sim 8\%$ level,  
while they are much larger, $\sim 12\%$, for \7li.  The reader may  
not notice that a band is also shown for \4he, since the uncertainty  
in Y is only at the $\sim 0.2\%$ level ($\sigma_{\rm Y} \approx  
0.0005$).  The results shown here are from the BBN code developed  
and refined over the years by my colleagues at The Ohio State  
University.  They are in excellent agreement with the published  
results of the Chicago group (Burles, Nollett \& Turner 2001) 
who, in a reanalysis of the relevant published cross sections  
have reduced the theoretical errors by roughly a factor of three 
for D and \3he and a factor of two for \7li.  The uncertainty in  
Y is largely due to the (very small) uncertainty in the neutron 
lifetime. 

The trends shown in Figure 1 are easy to understand based on our 
previous discussion.  D and \3he are burned to \4he.  The higher 
the nucleon density, the faster this occurs, leaving behind fewer 
nuclei of D or \3he.  The very slight increase of Y with $\eta$ 
is largely due to BBN starting earlier, at higher nucleon density 
(more complete burning of D, $^3$H, and \3he to \4he) and 
neutron-to-proton ratio (more neutrons, more \4he). The behavior 
of \7li is more interesting.  At relatively low values of $\eta 
~\la 3$, mass-7 is largely synthesized as \7li (by 
$^3$H($\alpha$,$\gamma$)\7li reactions) which is easily destroyed 
in collisons with protons.  So, as $\eta$ increases at low values, 
\7li/H decreases.  However, at relatively high values of $\eta ~\ga 
3$, mass-7 is largely synthesized as $^7$Be (via 
\3he($\alpha$,$\gamma$)$^7$Be reactions) which is more tightly 
bound than \7li and, therefore, harder to destroy.  As $\eta$ 
increases at high values, the abundance of $^7$Be increases.  
Later in the evolution of the universe, when it is cooler and 
neutral atoms begin to form, $^7$Be will capture an electron 
and $\beta$-decay to \7li. 
  
\subsection{Variations On A Theme: Non-Standard BBN} 
 
Before moving on, let's take a diversion to which we'll return 
again in \S5.  Suppose the standard model is modified through 
the addition of extra relativistic particles (\Deln $> 0$; SSG). 
Equivalently (ignoring some small differences), it could be 
that the gravitational constant in the early universe differs 
from its present value ($G \rightarrow G' \neq G$).  Depending 
on whether $G' > G$ or $G' < G$, the early universe expansion 
rate can be speeded up or slowed down compared to the standard 
rate.  For concreteness, let's assume that $S > 1$.  Now, there 
will be less time to destroy D and \3he, so their relic abundances 
will increase relative to the SBBN prediction.  There is less 
time for neutrons to transform to protons.  With more neutrons 
available, more \4he will be synthesized.  The changes in \7li 
are more complex.  At low $\eta$ there is less time to destroy 
\7li, so the relic \7li abundance increases.  At high $\eta$ 
there is less time to produce $^7$Be, so the relic \7li (mass-7) 
abundance decreases.

\begin{figure}[t!]\label{hevsdNnu234}  
\resizebox{\hsize}{!}{\includegraphics{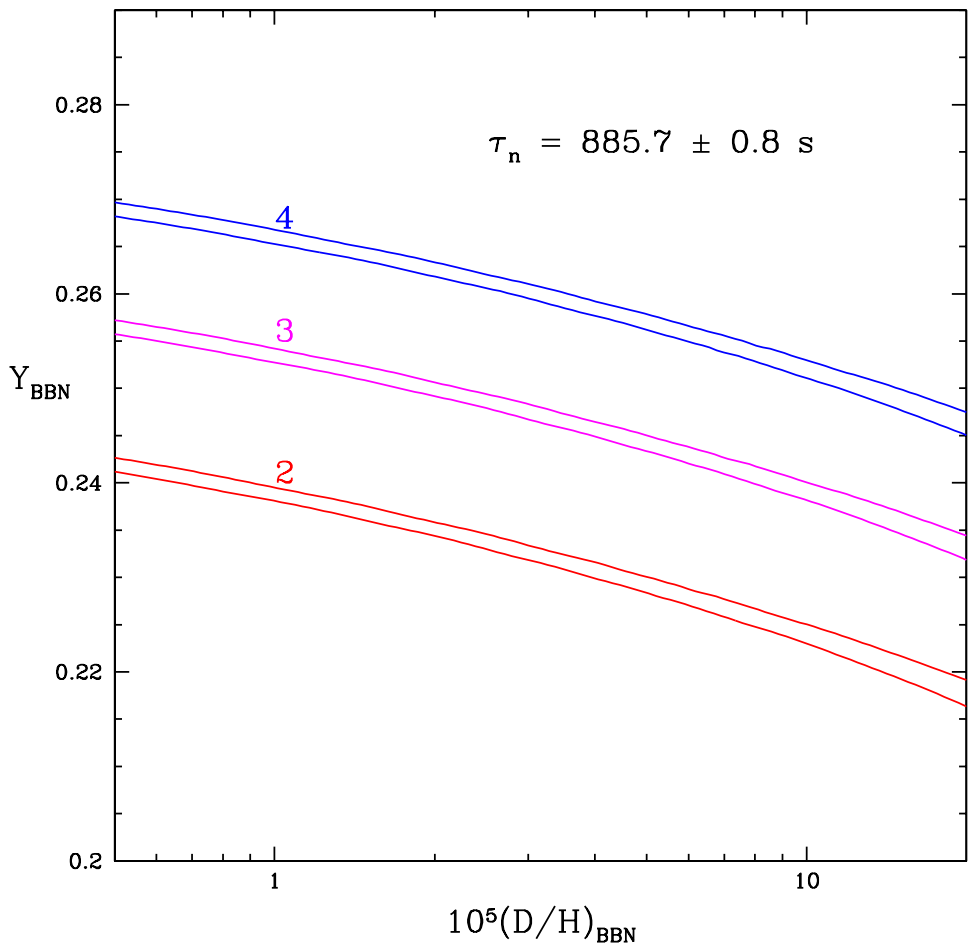}}  
\hfill  
\parbox[b]{\hsize}{  
\caption{The BBN-predicted primordial \4he mass fraction Y as a  
function of the BBN-predicted primordial Deuterium abundance  
(by number relative to Hydrogen) for three choices of N$_{\nu}$. 
The width of the bands represents the theoretical uncertainty, 
largely due to that of the neutron lifetime $\tau_{n}$.}}  
\end{figure}  
Since the \4he mass fraction is relatively insensitive to the 
baryon density, it provides an excellent probe of any changes  
in the expansion rate.  The faster the universe expands, the 
less time for neutrons to convert to protons, the more \4he 
will be synthesized.  The increase in Y for ``modest" changes 
in $S$ is roughly $\Delta $Y $ \approx 0.16(S-1) \approx 0.013 
\Delta N_{\nu}$.  
In Figure 2 are shown 
the BBN-predicted Y versus the BBN-predicted Deuterium abundance  
(relative to Hydrogen) for three choices of N$_{\nu}$ (N$_{\nu}  
\equiv 3 + \Delta N_{\nu}$). 
 
\section{Observational Status of the Relic Abundances} 
 
Armed with the SBBN-predicted primordial abundances, as well as 
with those in a variation on the standard model, we now turn to 
the observational data.  The four light nuclides of interest, D,  
\3he, \4he, and \7li follow different evolutionary paths in 
the post-BBN universe.  In addition, the observations leading 
to their abundance determinations are different for all four. 
Neutral D is observed in absorption in the UV; singly-ionized  
\3he is observed in emission in galactic \hii regions; both  
singly- and doubly-ionized \4he are observed in emission via 
their recombinations in extragalactic \hii regions; \7li is 
observed in absorption in the atmospheres of very metal-poor 
halo stars. The different histories and observational strategies 
provides some insurance that systematic errors affecting the 
inferred primordial abundances of any one of the light nuclides 
will not influence the inferred abundances of the others. 
 
\subsection{Deuterium} 
 
The post-BBN evolution of D is simple.  As gas is incorporated 
into stars the very loosely bound deuteron is burned to \3he 
(and beyond).  Any D which passes through a star is destroyed. 
Furthermore, there are no astrophysical sites where D can be 
produced in an abundance anywhere near that which is observed  
(Epstein, Lattimer, \& Schramm 1976).  As a result, as the 
universe evolves and gas is cycled through generations of 
stars, Deuterium is only destroyed.  Therefore, observations 
of the deuterium abundance anywhere, anytime, provide lower 
bounds on its primordial abundance.  Furthermore, if D can 
be observed in ``young" systems, in the sense of very little 
stellar processing, the observed abundance should be very close  
to the primordial value.  Thus, while there are extensive data 
on deuterium in the solar system and the local interstellar 
medium (ISM) of the Galaxy, it is the handful of observations of 
deuterium absorption in high-redshift (hi-$z$), low-metallicity  
(low-Z), QSO absorption-line systems (QSOALS) which are,  
potentially the most valuable.  In Figure 3 the extant data 
(circa November 2001) are shown for D/H as a function of 
redshift from the work of Burles \& Tytler (1998a,b), O'Meara 
\etal (2001), D'Odorico \etal (2001), and Pettini \& Bowen 
(2001).  Also shown for comparison are the local ISM D/H  
(Linsky \& Wood 2000) and that for the presolar nebula as  
inferred from solar system data (Geiss \& Gloeckler 1998, 
Gloeckler \& Geiss 2000). 
 
\begin{figure}[t!]\label{dvsz2}  
\resizebox{\hsize}{!}{\includegraphics{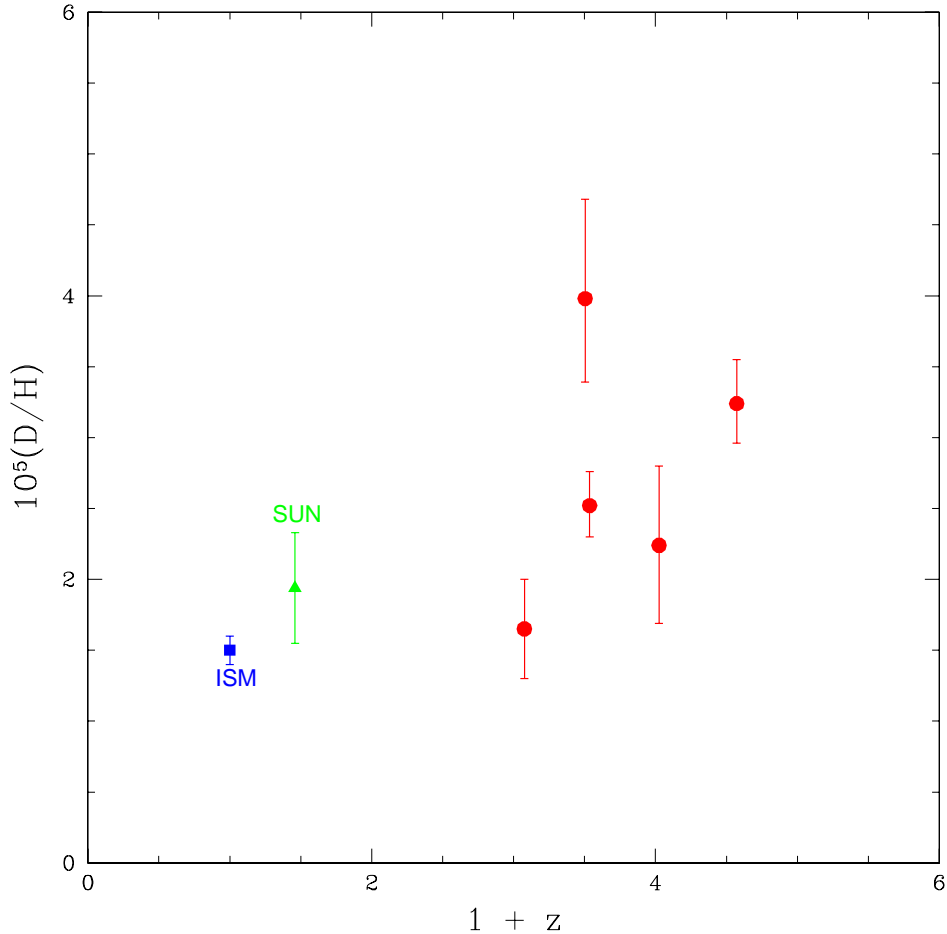}}  
\hfill  
\parbox[b]{\hsize}{  
\caption{The deuterium abundance, D/H, versus redshift, 
z, from observations of QSOALS (filled circles).  Also 
shown for comparison are the D-abundances for the local 
ISM (filled square) and the solar system (``Sun"; filled 
triangle).}}  
\end{figure}  
 
On the basis of our discussion of the post-BBN evolution 
of D/H, it would be expected that there should be a 
``Deuterium Plateau" at high redshift.  If, indeed, one 
is present, the dispersion in the limited set of current 
data hide it.  Alternatively, to explore the possibility 
that the D-abundances may be correlated with the metallicity 
of the QSOALS, we may plot the observed D/H versus the 
metallicity, as measured by [Si/H], for these absorbers.  
This is shown in Figure 4 where there is some evidence for 
an (unexpected!) increase in D/H with decreasing [Si/H]; 
once again, the dispersion in D/H hides any plateau. 
 
\begin{figure}[t!]\label{dvssi2}  
\resizebox{\hsize}{!}{\includegraphics{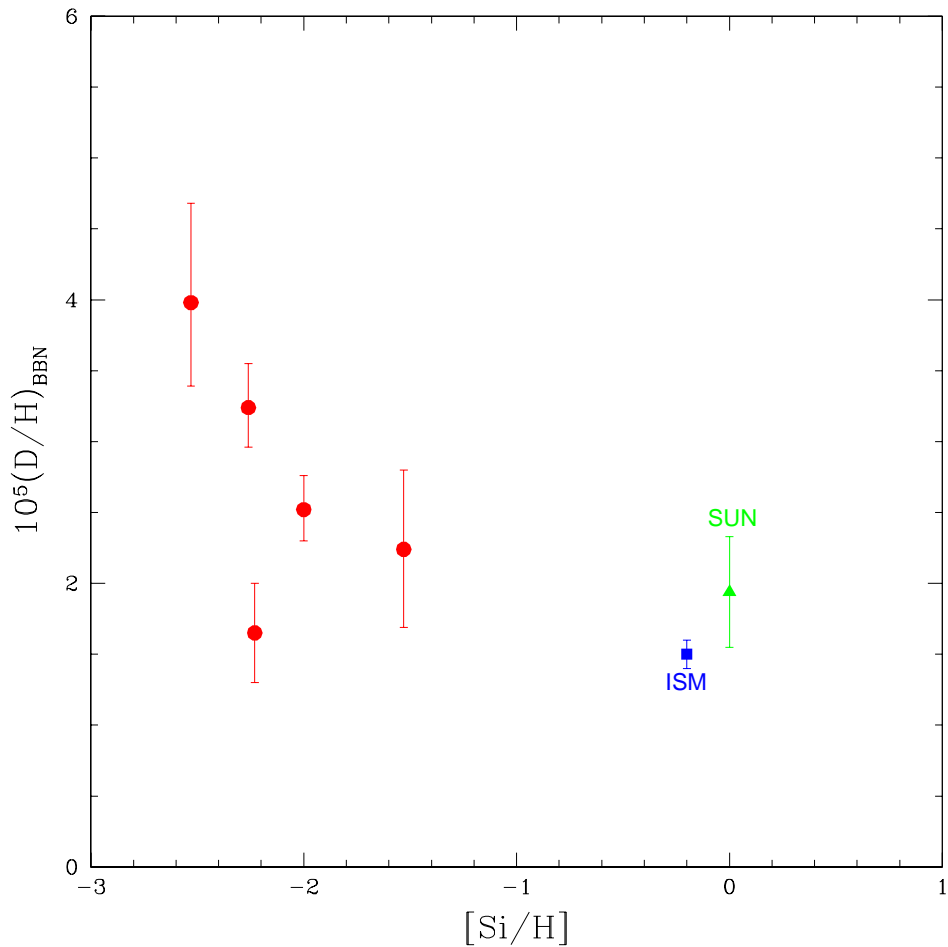}}  
\hfill  
 \parbox[b]{\hsize}{  
\caption{The deuterium abundance, D/H, versus metallicity 
([Si/H]) for the same QSOALS as in Figure 3 (filled circles).  
Also shown for comparison are the D-abundances for the local 
ISM (filled square) and the solar system (``Sun"; filled 
triangle).}}  
\end{figure}  

Aside from observational errors, there are several sources of  
systematic error which may account for the observed dispersion. 
For example, the Ly$\alpha$ absorption of \h1 in these systems 
is saturated, potentially hiding complex velocity structure. 
Usually, but not always, this velocity structure can be revealed 
in the higher lines of the Lyman series and, expecially, in the 
narrower metal-absorption lines.  Recall, also, that the lines 
in the Lyman series of \d1 are identical to those of \h1, only 
shifted by $\approx$ 81 km/s.  Given the highly saturated \h1 Ly$ 
\alpha$, it may be difficult to identify which, and how much,  
of the \h1 corresponds to an absorption feature identified as 
\d1 Ly$\alpha$.  Furthermore, are such features really \d1 or, 
an interloping, low column density \h1-absorber?  After all, 
there are many more low-, rather than high-column density \h1 
systems.  Statistically, the highest column density absorbers 
may be more immune to these systematic errors.  Therefore, in 
Figure 5 are shown the very same D/H data, now plotted against 
the neutral hydrogen column density.  The three  highest column 
density absorbers (Damped Ly$\alpha$ Absorbers: DLAs) fail to 
reveal the D-plateau, although it may be the case that {\it some} 
of the D associated with the two lower column density systems 
may be attributable to an interloper, which would reduce the 
D/H inferred for them. 
 
\begin{figure}[t!]\label{dvsh2}  
\resizebox{\hsize}{!}{\includegraphics{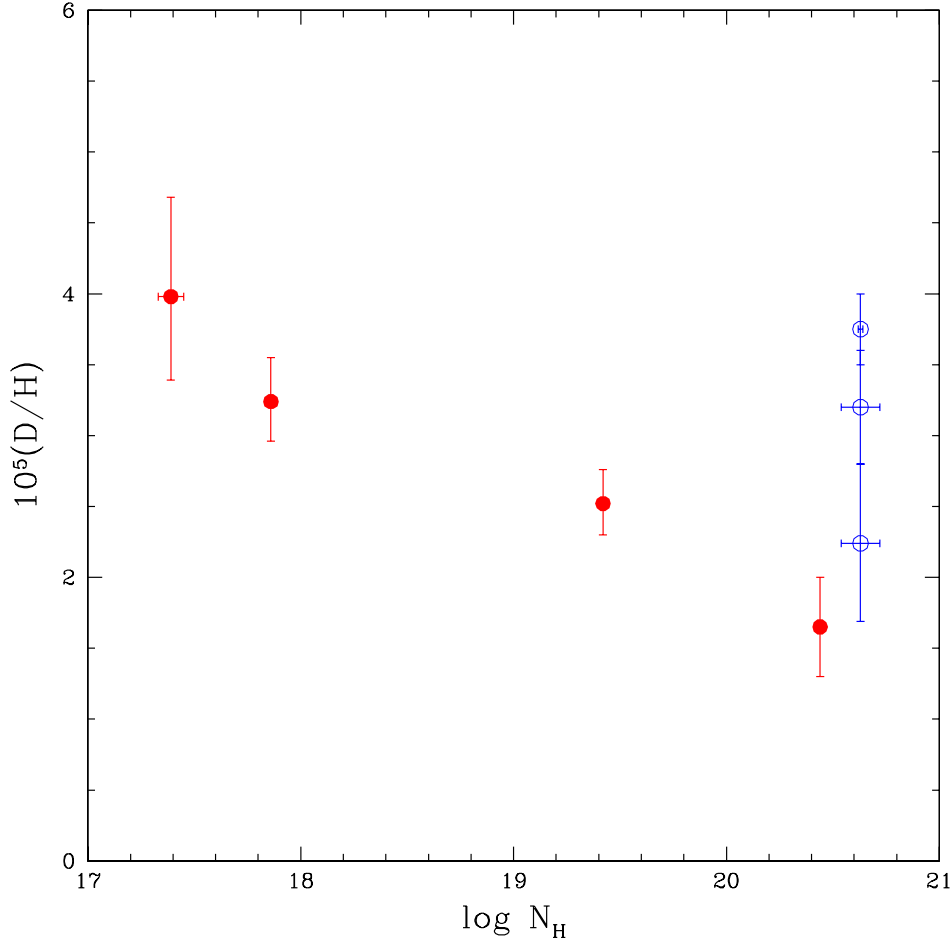}}  
\hfill  
\parbox[b]{\hsize}{  
\caption{The same QSOALS D/H data as in Figures 3 \& 4 versus 
the \h1 column density (log scale).  The open circle symbols
are for the original (D'Odorico \etal 2001) and the revised 
(Levshakov \etal 2002) D/H values for Q0347-3819.}}  
\end{figure}  
  
Actually, the situation is even more confused.  The highest 
column density absorber, from D'Odorico \etal (2001), was 
reobserved by Levshakov \etal (2002) and revealed to have a 
more complex velocity structure.  As a result, the D/H has 
been revised from $2.24 \times 10^{-5}$ to $3.2 \times 10^{-5}$ 
to $3.75 \times 10^{-5}$. To this theorist, at least, this 
evolution suggests that the complex velocity structure in 
this absorber renders it suspect for determining primordial 
D/H.  The sharp-eyed reader may notice that if this D/H 
determination if removed from Figure 5, there is a hint of 
an {\it anticorrelation} between D/H and N$_{\rm H}$ among 
the remaining data points, suggesting that interlopers may 
be contributing to (but not necessarily dominating) the 
inferred \d1 column density. 

However, the next highest column density \h1-absorber (Pettini 
\& Bowen 2001), has the lowest D/H ratio, at a value 
indistinguishable from the ISM and solar system abundances.  
Why such a high-$z$, low-Z system should have destroyed so 
much of its primordial D so early in the evolution of the 
universe, apparently without  producing very many heavy elements, 
is a mystery.  If, for no really justifiable reason, this system 
is arbitrarily set aside, only the three ``UCSD" systems of Burles 
\& Tytler (1998a,b) and O'Meara \etal (2001) remain.  The weighted 
mean for these three absorbers is D/H $ = 3.0 \times 10^{-5}$.  
O'Meara \etal note the larger than expected dispersion, even 
for this subset of D-abundances, and they suggest increasing 
the formal error in the mean, leading to: (D/H)$_{\rm P} = 
3.0 \pm 0.4 \times 10^{-5}$.  I will be even more cautious; 
when the SBBN predictions are compared with the primordial 
abundances inferred from the data, I will adopt: (D/H)$_{\rm P} 
= 3.0 ^{+1.0}_{-0.5} \times 10^{-5}$.  Since the primordial D 
abundance is sensitive to the baryon abundance (D/H $\propto 
\eta^{-1.6}$), even these perhaps overly generous errors will 
still result in SBBN-derived baryon abundances which are 
accurate to 10 -- 20\%. 
 
\subsection{Helium-3} 
 
The post-BBN evolution of \3he is much more complex than that 
of D.  Indeed, when D is incorporated into a star it is rapidly 
burned to \3he, increasing the \3he abundance.  The more tightly 
bound \3he, with a larger coulomb barrier, is more robust than 
D to nuclear burning.  Nonetheless, in the hotter interiors of 
most stars \3he is burned to \4he and beyond.  However, in the 
cooler, outer layers of most stars, and throughout most of the 
volume of the cooler, lower mass stars, \3he is preserved 
(Iben 1967, Rood, Steigman, \& Tinsley 1976; Iben \& Truran 
1978, Dearborn, Schramm, \& Steigman 1986; Dearborn, Steigman, 
\& Tosi 1996). As a result, prestellar \3he is enhanced by the 
burning of prestellar D, and some, but not all, of this \3he 
survives further stellar processing.  However, there's more to 
the story.  As stars burn hydrogen to helium and beyond, some 
of their newly synthesized \3he will avoid further processing 
so that the cooler, lower mass stars should be significant 
post-BBN sources of \3he. 

Aside from studies of meteorites and in samples of the lunar 
soil (Reeves \etal 1973, Geiss \& Gloeckler 1998, Gloeckler 
\& Geiss 2000), \3he is only observed via its hyperfine line 
(of singly-ionized \3he) in interstellar \hii regions in the 
Galaxy.  It is, therefore, unavoidable that models of stellar 
yields and Galactic chemical evolution are required in order 
to go from here and now (ISM) to there and then (BBN).  It 
has been clear since the early work of Rood, Steigman and 
Tinsley (1976) that according to such models, \3he should 
have increased from the big bang and, indeed, since the 
formation of the solar system (see, \eg Dearborn, Steigman 
\& Tosi 1996 and further references therein).  For an element 
whose abundance increases with stellar processing, there should 
also be a clear gradient in abundance with galactocentric 
distance.  Neither of these expectations is borne out by 
the data (Rood, Bania \& Wilson 1992; Balser \etal 1994, 
1997, 1999; Bania, Rood \& Balser 2002) which shows no 
increase from the time of the formation of the solar system, 
nor any gradient within the Galaxy.  The most likely explanation 
is that before the low mass stars can return their newly 
processed \3he to the interstellar medium, it is mixed to 
the hotter interior and destroyed (Charbonnel 1995, Hogan 
1995).  Whatever the explanation, the data suggest that 
for \3he there is a delicate balance between production 
and destruction.  As a result, the model-dependent uncertainties 
in extrapolating from the present data to the primordial 
abundances are large, limiting the value of \3he as a 
baryometer.  For this reason I will not dwell further on 
\3he in these lectures; for further discussion and references, 
the interested reader is referred to the excellent review 
by Tosi (2000).  

\subsection{Helium-4} 
 
\begin{figure}[t!]\label{keith}  
\resizebox{\hsize}{!}{\includegraphics{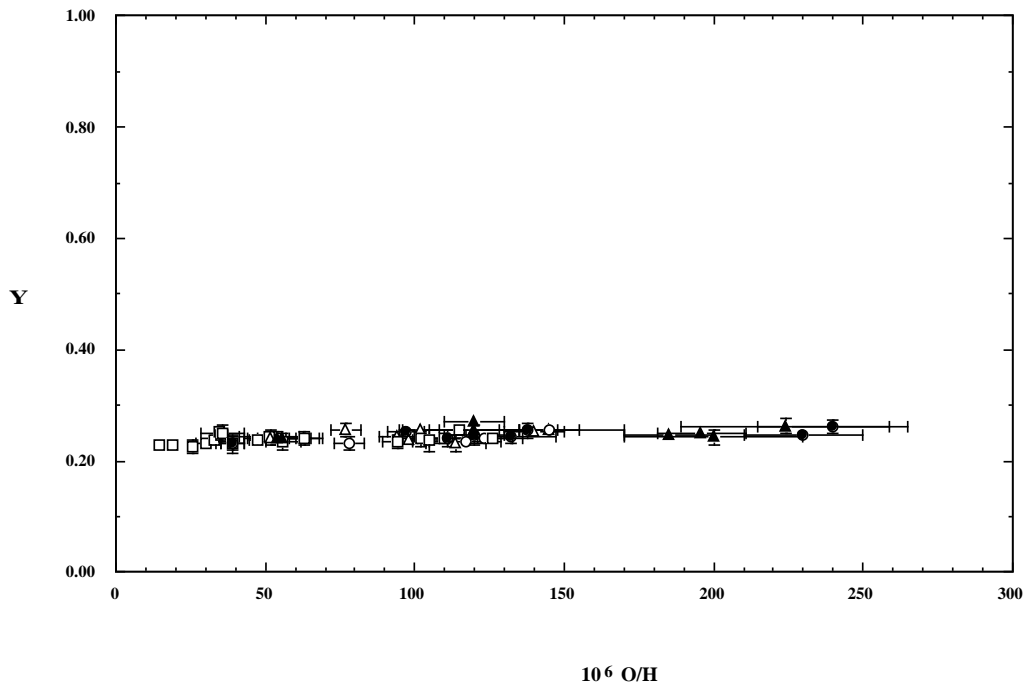}}  
\hfill  
\parbox[b]{\hsize}{  
\caption{The \4he mass fraction, Y, inferred from 
observations of low-metallicity, extragalactic \hii 
regions versus the oxygen abundance in those regions.}}  
\end{figure}  
  
Helium-4 is the second most abundant nuclide in the universe, 
after hydrogen.  In the post-BBN epochs the net effect of gas 
cycling though generations of stars is to burn hydrogen to 
helium, increasing the \4he abundance.  As with deuterium, a 
\4he ``plateau" is expected at low metallicity.  Although \4he 
is observed in the Sun and in Galactic \hii regions, the most 
relevant data for inferring its primordial abundance (the 
plateau value) is from observations of the helium and hydrogen  
recombination lines in low-metallicity, extragalactic \hii  
regions.  The present inventory of such observations is  
approaching of order 100.  It is, therefore, not surprising  
that even with modest observational errors for any individual  
\hii region, the statistical uncertainty in the inferred  
primordial abundance may be quite small.  Especially in  
this situation, care must be taken with hitherto ignored  
or unaccounted for corrections and systematic errors. 
 
\begin{figure}[t!]\label{pplvsit}  
\resizebox{\hsize}{!}{\includegraphics{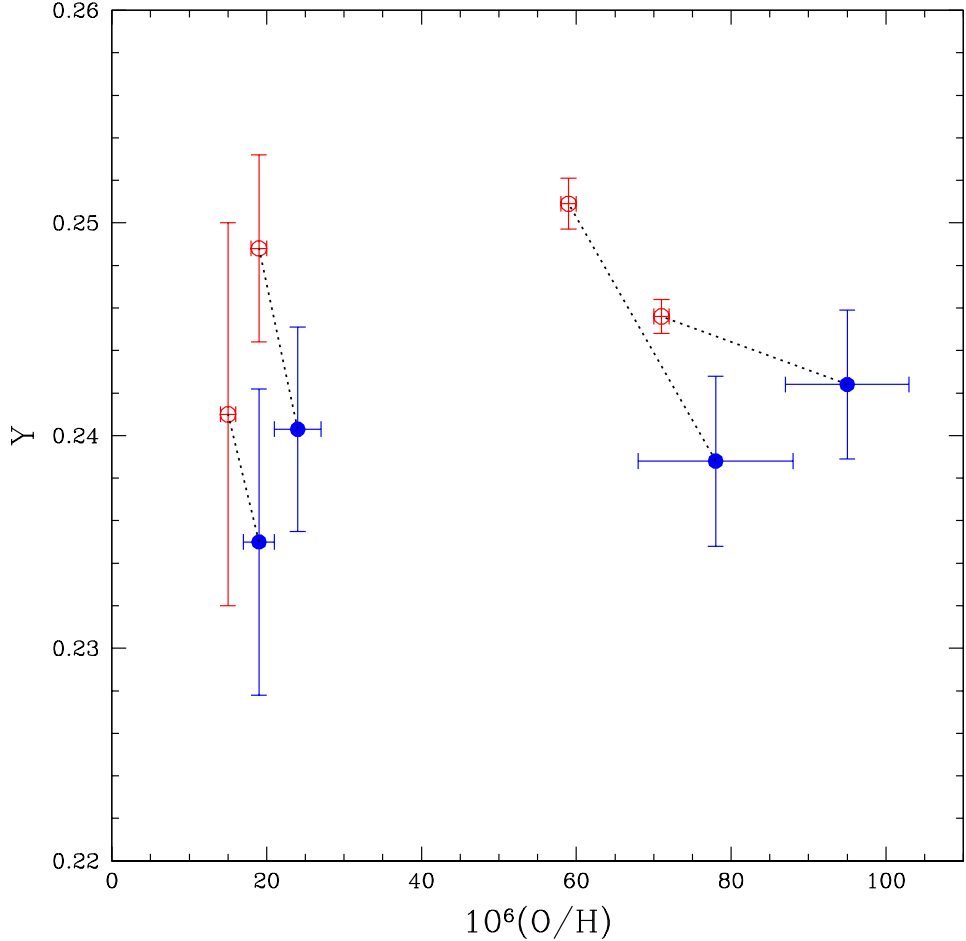}}  
\hfill  
\parbox[b]{\hsize}{  
\caption{The Peimbert, Peimbert, \& Luridiana (2002) 
reanalysis of the Helium-4 abundance data for 4 of 
the IT \hii regions.  The open circles are the IT 
abundances, while the filled circles are from PPL.}}  
\end{figure}  
  
In Figure 6 is shown a compilation of the data used by  
Olive \& Steigman (1995) and Olive, Skillman, \& Steigman  
(1997), along with the independent data from Izotov, Thuan,  
\& Lipovetsky (1997) and Izotov \& Thuan (1998).  To track  
the evolution of the \4he mass fraction, Y is plotted versus  
the \hii region oxygen abundance.  These \hii regions are  
all metal-poor, ranging from $\sim 1/2$ down to $\sim 1/30$  
of solar (for a solar oxygen abundance of O/H $ = 5 \times  
10^{-4}$; Allende-Prieto, Lambert, \& Asplund 2001).  A key 
feature of Figure 6 is that independent of whether there is 
a statistically significant non-zero slope to the Y vs. O/H 
relation, there is a \4he plateau!  Since Y is increasing 
with metallicity, the relic abundance can either be bounded 
from above by the lowest metallicity regions, or the Y vs. 
O/H relation determined observationally may be extrapolated 
to zero metallicity (a not very large extrapolation, $\Delta$Y 
$ \approx -0.001$). 

The good news is that the data reveal a well-defined 
primordial abundance for \4he.  The bad news is that 
the scale of Figure 6 hides the very small statistical 
errors, along with a dichotomy between the OS/OSS and 
ITL/IT primordial helium abundance determinations 
(Y$_{\rm P}$(OS) $= 0.234 \pm 0.003$ versus Y$_{\rm P}$(IT) 
$= 0.244 \pm 0.002$).  Furthermore, even if one adopts 
the IT/ITL data, there are corrections which should be 
applied which change the inferred primordial \4he abundance 
by more than their quoted statistical errors (see, \eg 
Steigman, Viegas \& Gruenwald 1997; Viegas, Gruenwald 
\& Steigman 2000; Sauer \& Jedamzik 2002, Gruenwald, 
Viegas \& Steigman 2002(GSV); Peimbert, Peimbert \& 
Luridiana 2002).  In recent high quality observations 
of a relatively metal-rich SMC \hii region, Peimbert, 
Peimbert and Ruiz (2000; PPR) derive Y$_{\rm SMC} = 
0.2405 \pm 0.0018$.  This is already {\it lower} than 
the IT-inferred {\it primordial} \4he abundance.  Further, 
when PPR extrapolate this abundance to zero-metallicity, 
they derive Y$_{\rm P}({\rm PPR}) = 0.2345 \pm 0.0026$, 
lending some indirect support for the lower OS/OSS value.  

Recently, Peimbert, Peimbert, \& Luridiana (2002; PPL)  
have reanalyzed the data from four of the IT \hii regions.   
When correcting for the \hii region temperatures and  
the temperature fluctuations, PPL derive systematically  
lower helium abundances as shown in Figure 7.  PPL also 
combine their redetermined abundances for these four 
\hii regions with the recent accurate determination of 
Y in the more metal-rich SMC \hii region (PPR).  These 
five data points are consistent with zero slope in the 
Y vs. O/H relation, leading to a primordial abundance 
Y$_{\rm P} = 0.240 \pm 0.001$.  However, this very 
limited data set is also consistent with $\Delta$Y $ 
\approx 40($O/H).  In this case, the extrapolation to 
zero metallicity, starting at the higher SMC metallicity, 
leads to the considerably smaller estimate of Y$_{\rm P} 
\approx 0.237$. 
 
It seems clear that until new data address the unresolved 
systematic errors afflicting the derivation of the primordial 
helium abundance, the true errors must be much larger than 
the statistical uncertainties.  For the comparisons between 
the predictions of SBBN and the observational data to be 
made in the next section, I will adopt the Olive, Steigman 
\& Walker (2000; OSW) compromise: Y$_{\rm P} = 0.238 \pm 
0.005$; the inflated errors are an attempt to account for 
the poorly-constrained systematic uncertaintiess. 
 
\subsection{Lithium-7}

Lithium-7 is fragile, burning in stars at a relatively low
temperature.  As a result, the majority of interstellar 
\7li cycled through stars is destroyed.  For the same
reason, it is difficult for stars to create new \7li and
return it to the ISM before it is destroyed by nuclear
burning. As the data in Figure 8 reveal, only relatively
late in the evolution of the Galaxy, when the metallicity
approaches solar, does the lithium abundance increase
noticeably.  However, the intermediate-mass nuclides $^6$Li, 
\7li, $^9$Be, $^{10}$B, and $^{11}$B can be synthesized via 
Cosmic Ray Nucleosynthesis (CRN), either by alpha-alpha fusion 
reactions, or by spallation reactions (nuclear breakup) between 
protons and alpha particles on the one hand and CNO nuclei on 
the other.  In the early Galaxy, when the metallicity is low, 
the post-BBN production of lithium is expected to be subdominant 
to the pregalactic, BBN abundance.  This is confirmed in Figure 
8 by the ``Spite Plateau'' (Spite \& Spite 1982), the absence 
of a significant slope in the Li/H versus [Fe/H] relation 
at low metallicity.  This plateau is a clear signal of the 
primordial origin of the low-metallicity lithium abundance. 
Notice, also, the enormous {\it spread} among the lithium 
abundances at higher metallicity.  This range in Li/H results
from the destruction/dilution of lithium on the surfaces of
the observed stars, implying that it is the {\it upper envelope}
of the Li/H versus [Fe/H] relation which preserves the history
of the Galactic lithium evolution.  Note, also, that at low 
metallicity this dispersion is much narrower, suggesting that 
the corrections for depletion/dilution are much smaller for 
the Pop II stars.

\begin{figure}[t!]\label{spiteplateau}  
\resizebox{\hsize}{!}{\includegraphics{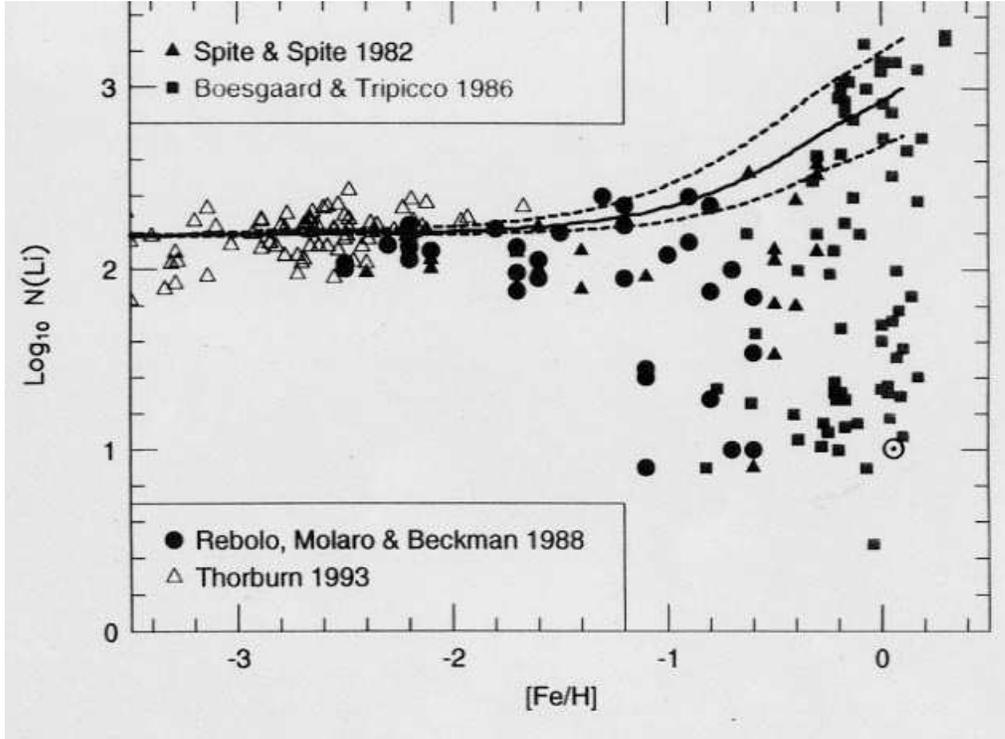}}  
\hfill  
\parbox[b]{\hsize}{
\caption{A compilation of the lithium abundance 
data from stellar observations as a function of 
metallicity.  N(Li) $\equiv 10^{12}$(Li/H) and 
[Fe/H] is the usual metallicity relative to solar.  
Note the ``Spite Plateau" in Li/H for [Fe/H] 
$\la -2$.}}  
\end{figure}  
 
As with the other relic nuclides, the dominant uncertainties 
in estimating the primordial abundance of \7li are not
statistical, they are systematic.  Lithium is observed in 
the atmospheres of cool stars (see Lambert (2002) in these 
lectures).  It is the metal-poor, Pop II halo stars that 
are of direct relevance for the BBN abundance of \7li.  
Uncertainties in the lithium equivalent width measurements, 
in the temperature scales for these Pop II stars, and in 
their model atmospheres, dominate the overall error budget.  
For example, Ryan \etal (2000), using the Ryan, Norris \& 
Beers (1999) data, infer [Li]$_{\rm P} \equiv 12 + $log(Li/H)$ 
= 2.1$, while Bonifacio \& Molaro (1997) and Bonifacio,
Molaro \& Pasquini (1997) derive [Li]$_{\rm P} = 2.2$,
and Thorburn (1994) finds [Li]$_{\rm P} = 2.3$.  From
recent observations of stars in a metal-poor globular
cluster, Bonifacio \etal (2002) derive [Li]$_{\rm P}
= 2.34 \pm 0.056$.  But, there's more.

The very metal-poor halo stars used to define the
lithium plateau are very old.  They have had the
most time to disturb the prestellar lithium which
may survive in their cooler, outer layers.  Mixing
of these outer layers with the hotter interior where
lithium has been destroyed will dilute the surface
abundance.  Pinsonneault \etal (1999, 2002) have
shown that rotational mixing may decrease the
surface abundance of lithium in these Pop II stars
by 0.1 -- 0.3 dex while maintaining a rather narrow
{\it dispersion} among their abundances (see also,
Chaboyer \etal 1992; Theado \& Vauclair 2001, Salaris 
\& Weiss 2002).

In Pinsonneault \etal (2002) we adopted for our
baseline (Spite Plateau) estimate [Li]$ = 2.2 \pm 0.1$; 
for an overall depletion factor we chose 0.2 $\pm 0.1$ 
dex.  Combining these {\it linearly}, we derived an 
estimate of the primordial lithium abundance of
[Li]$_{\rm P} = 2.4 \pm 0.2$.  I will use this in
the comparison between theory and observation to
be addressed next.
  
\section{Confrontation Of Theoretical Predictions With 
Observational Data} 

As the discussion in the previous section should have made
clear, the attempts to use a variety of observational data
to infer the BBN abundances of the light nuclides is fraught
with evolutionary uncertainties and dominated by systematic
errors.  It may be folly to represent such data by a ``best'' 
value along with normally distributed errors.  Nonetheless, 
in the absence of a better alternative, this is what will 
be done in the following.  

\begin{itemize}

\item{Deuterium}

From their data along the lines-of-sight to three QSOALS, 
O'Meara \etal (2001) recommend (D/H)$_{\rm P} = 3.0 \pm 
0.4 \times 10^{-5}$.  While I agree this is likely a good 
estimate for the {\it central} value, the spread among the
extant data (see Figures 3 -- 5) favors a larger uncertainty.
Since D is only destroyed in the post-BBN universe, the solar 
system and ISM abundances set a floor to the primordial value.  
Keeping this in mind, I will adopt asymmetric errors ($\sim 
1\sigma$): (D/H)$_{\rm P} = 3.0 ^{+1.0}_{-0.5} \times 10^{-5}$.

\item{Helium-4}

In our discussion of \4he as derived from hydrogen and
helium recombination lines in low-metallicity, extragalactic
\hii regions it was noted that the inferred primordial
mass fraction varied from Y$_{\rm P} = 0.234 \pm 0.003$ 
(OS/OSS), to Y$_{\rm P} = 0.238 \pm 0.003$ (PPL and GSV),
to Y$_{\rm P} = 0.244 \pm 0.002$ (IT/ITL).  Following the
recommendation of OSW, here I will choose as a compromise
Y$_{\rm P} = 0.238 \pm 0.005$.

\item{Lithium-7}

Here, too, the spread in the level of the ``Spite Plateau''
dominates the formal errors in the means among the different
data sets.  To this must be added the uncertainties due to
temperature scale and model atmospheres, as well as some
allowance for dilution or depletion over the long lifetimes
of the metal-poor halo stars. Attempting to accomodate 
all these sources of systematic uncertainty, I adopt 
the Pinsonneault \etal (2002) choice of [Li]$_{\rm P} 
= 2.4 \pm 0.2$.
\end{itemize}

As discussed earlier, the stellar and Galactic chemical 
evolution uncertainties afflicting \3he are so large as 
to render the use of \3he to probe or test BBN problematic;
therefore, I will ignore \3he in the subsequent discussion.  
There are a variety of equally valid approaches to using 
D, \4he, and \7li to test and constrain the standard models 
of cosmology and particle physics (SBBN).  In the approach 
adopted here deuterium will be used to constrain the baryon 
density ($\eta$ or, equivalently, \obh).  Within SBBN, this 
leads to predictions of \Yp and [Li]$_{\rm P}$.  Indeed, 
once the primordial deuterium abundance is chosen, $\eta$ 
may be eliminated and both \Yp and [Li]$_{\rm P}$ predicted 
directly, thereby testing the consistency of SBBN.
 
\begin{figure}[t!]\label{dvseta}  
\resizebox{\hsize}{!}{\includegraphics{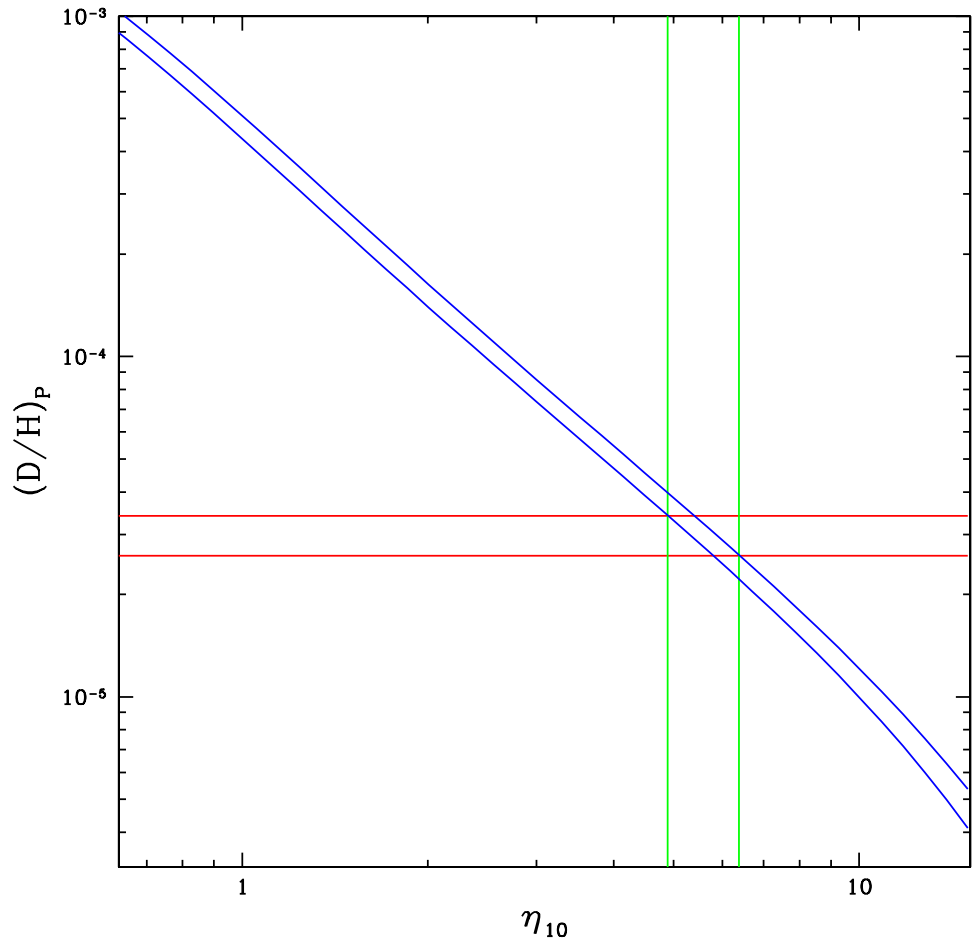}}  
\hfill  
\parbox[b]{\hsize}{
\caption{The diagonal band is the SBBN-predicted deuterium
abundance (by number relative to hydrogen) as a function of
the nucleon-to-photon ratio $\eta_{10}$ (the width of the 
band accounts for the theoretical uncertainties in the SBBN
prediction).  The horizontal band is the $\pm 1\sigma$ range 
in the adopted primordial deuterium abundance.  The vertical 
band is, approximately, the corresponding SBBN-predicted 
$\eta$ range.}}  
\end{figure}  

\subsection{Deuterium -- The Baryometer Of Choice}

Recall that D is produced (in an astrophysically interesting 
abundance) {\bf ONLY} during BBN.  The predicted primordial 
abundance is sensitive to the baryon density (D/H ~$\propto 
\eta^{-1.6}$).  Furthermore, during post-BBN evolution, as 
gas is cycled through stars, deuterium is {\bf ONLY} destroyed, 
so that for the ``true'' abundance of D anywhere, at any time, 
(D/H)$_{\rm P} \ge ({\rm D}/{\rm H})_{\rm TRUE}$.  That's the 
good news.  The bad news is that the spectra of \h1 and \d1 
are identical, except for the wavelength/velocity shift in 
their spectral lines.  As a result, the true D-abundance may 
differ from that inferred from the observations if {\it some} 
of the presumed \d1 is actually an \h1 interloper masquerading 
as \d1: (D/H)$_{\rm TRUE} \le ({\rm D}/{\rm H})_{\rm OBS}$.  
Because of these opposing effects, the connection between 
(D/H)$_{\rm OBS}$ and (D/H)$_{\rm P}$ is not predetermined; 
the data themselves which must tell us how to relate the two.  
With this caveat in mind, it will be assumed that the value 
of (D/H)$_{\rm P}$ identified above is a fair estimate of the 
primordial D abundance.  Using it, the SBBN-predicted baryon 
abundance may be determined.  The result of this comparison 
is shown in Figure 9 where, approximately, the overlap between 
the SBBN-predicted band and that from the data fix the allowed 
range of $\eta$. 

\subsection{SBBN Baryon Density -- The Baryon Density 
At 20 Minutes}

The universal abundance of baryons which follows from
SBBN and our adopted primordial D-abundance is: $\eta_
{10} = 5.6 ^{+0.6}_{-1.2}$ (\obh $= 0.020 ^{+0.002}_
{-0.004}$).  For the HST Key Project recommended value 
for $H_{0}$ ($h = 0.72 \pm 0.08$; Freedman \etal 2001), 
the fraction of the present universe critical density 
contributed by baryons is small, $\Omega_{\rm B} \approx 
0.04$.  In Figure 10 is shown a comparison among the
various determinations of the present mass/energy density 
(as a fraction of the critical density), baryonic as well 
as non-baryonic. It is clear from Figure 10 that the
present universe ($z ~\la 1$) baryon density inferred from 
SBBN far exceeds that inferred from emission/absorption 
observations (Persic \& Salucci 1992, Fukugita, Hogan 
\& Peebles 1998).  The gap between the upper bound to
luminous baryons and the BBN band is the ``dark baryon
problem'': at present, most of the baryons in the 
universe are dark.  Evidence that although dark, the
baryons are, indeed, present comes from the absorption
observed in the Ly$\alpha$ forest at redshifts $z \approx 
2 - 3$ (see, \eg Weinberg \etal 1997).  The gap between
the BBN band and the band labelled by $\Omega_{\rm M}$
is the ``dark matter problem'': the mass density inferred
from the structure and movements of the galaxies and
galaxy clusters far exceeds the SBBN baryon contribution.
Most of the mass in the universe must be nonbaryonic.
Finally, the gap from the top of the $\Omega_{\rm M}$ 
band to $\Omega = 1$ is the ``dark energy problem''.

\begin{figure}[t!]\label{omegab}  
\resizebox{\hsize}{!}{\includegraphics{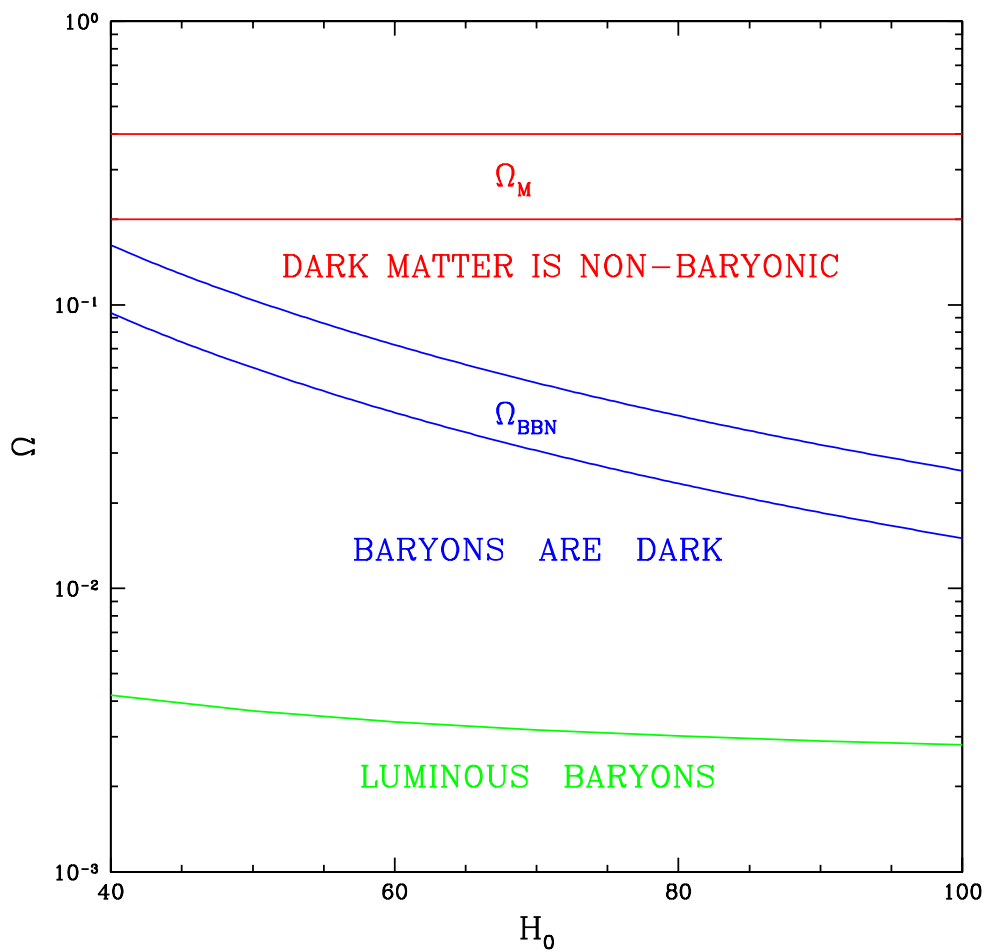}}  
\hfill  
\parbox[b]{\hsize}{
\caption{The various contributions to the present 
universal mass/energy density, as a fraction of 
the critical density ($\Omega$), as a function of 
the Hubble parameter ($H_0$). The curve labelled 
Luminous Baryons is an estimate of the upper bound 
to those baryons seen at present ($z ~\la 1$) either 
in emission or absorption (see the text).  The band 
labelled BBN represents the D-predicted SBBN baryon 
density.  The band labelled by ``M'' ($\Omega_{\rm M} 
= 0.3 \pm 0.1$) is an estimate of the current mass 
density in nonrelativistic particles (``Dark Matter'').}}  
\end{figure}  

\subsection{CMB Baryon Density -- The Baryon Density 
At A Few Hundred Thousand Years}

As discussed in the first lecture, the early universe 
is hot and dominated by relativistic particles 
(``radiation'').  As the universe expands and cools, 
nonrelativistic particles (``matter'') come to dominate 
after a few hundred thousand years, and any preexisting 
density perturbations can begin to grow under the 
influence of gravity.  On length scales determined by 
the density of baryons, oscillations (``sound waves'') 
in the baryon-photon fluid develop.  At a redshift of 
$z \sim 1100$ the electron-proton plasma combines 
(``recombination) to form neutral hydrogen which is 
transparent to the CMB photons.  Free to travel throughout 
the post-recombination universe, these CMB photons preserve 
the record of the baryon-photon oscillations as small 
temperature fluctuations in the CMB spectrum.  Utilizing 
recent CMB observations (Lee \etal 2001; Netterfield \etal 
2002; Halverson \etal 2002), many groups have inferred the 
intermediate age universe baryon density.  The work of our 
group at OSU (Kneller \etal 2002) is consistent with more 
detailed analyses and is the one I adopt for the purpose 
of comparison with the SBBN result: $\eta_{10} = 6.0 \pm 
0.6$; \obh $= 0.022 \pm 0.002$.

\subsection{The Baryon Density At 10 Gyr}

Although the majority of baryons in the recent/present
universe are dark, it is still possible to constrain 
the baryon density indirectly using observational data
(see, \eg Steigman, Hata \& Felten 1999, Steigman, Walker 
\& Zentner 2000; Steigman 2001).  The magnitude-redshift
relation determined by observations of type Ia supernovae
(SNIa) constrain the relation between the present matter
density ($\Omega_{\rm M}$) and that in a cosmological
constant ($\Omega_{\Lambda}$).  The allowed region in
the $\Omega_{\Lambda}$ -- $\Omega_{\rm M}$ plane derived
from the observations of Perlmutter \etal (1997), Schmidt
\etal (1998), and Perlmutter \etal (1999) are shown in
Figure 11. 
 
\begin{figure}[t!]\label{sn1a}  
\resizebox{\hsize}{!}{\includegraphics{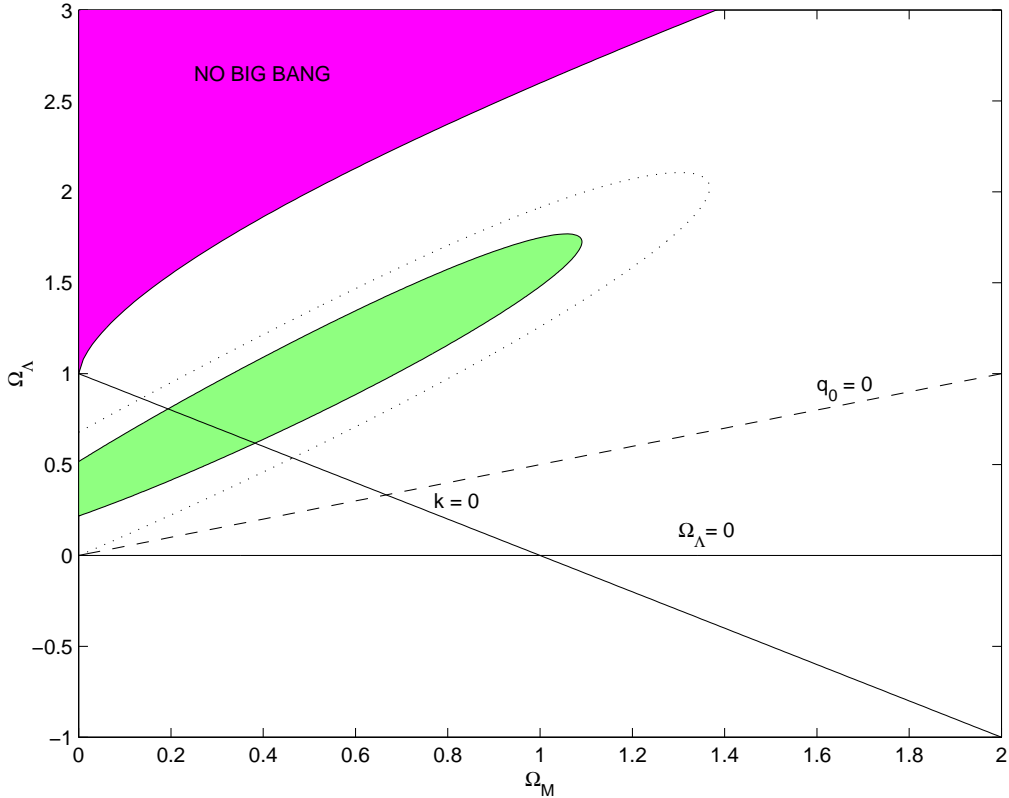}}  
\hfill  
\parbox[b]{\hsize}{
\caption{
The 68\% (solid) and 95\% (dotted) 
contours in the $\Omega_{\Lambda}$ -- $\Omega_{\rm M}$ 
plane consistent with the SNIa data (see the text).
Geometrically flat models lie along the line labelled 
k = 0.}}  
\end{figure}

If, in addition, it is {\it assumed} that the universe 
is flat ($\kappa = 0$; an assumption supported by the 
CMB data), a reasonably accurate determination of 
$\Omega_{\rm M}$ results: $\Omega_{\rm M}({\rm SNIa;Flat}) 
= 0.28 ^{+0.08}_{-0.07}$ (Steigman, Walker \& Zentner 
2000; Steigman 2001).  But, how to go from the matter 
density to the baryon density?  For this we utilize rich 
clusters of galaxies, the largest collapsed objects, 
which provide an ideal probe of the baryon {\it fraction}
in the present universe $f_{\rm B}$.  X-ray observations
of the hot gas in clusters, when corrected for the baryons
in stars (albeit not for any dark cluster baryons), can
be used to estimate $f_{\rm B}$.  Using the Grego \etal 
(2001) observations of the Sunyaev-Zeldovich effect in
clusters, Steigman, Kneller \& Zentner (2002) estimate
$f_{\rm B}$ and derive a present-universe ($t_0 \approx 
10$~Gyr; $z ~\la 1$) baryon density: $\eta_{10} = 5.1 
^{+1.8}_{-1.4}$ (\obh $= 0.019 ^{+0.007}_{-0.005}$).

\subsection{Baryon Density Concordance}

\begin{figure}[t!]\label{b-likelihoods}  
\resizebox{\hsize}{!}{\includegraphics{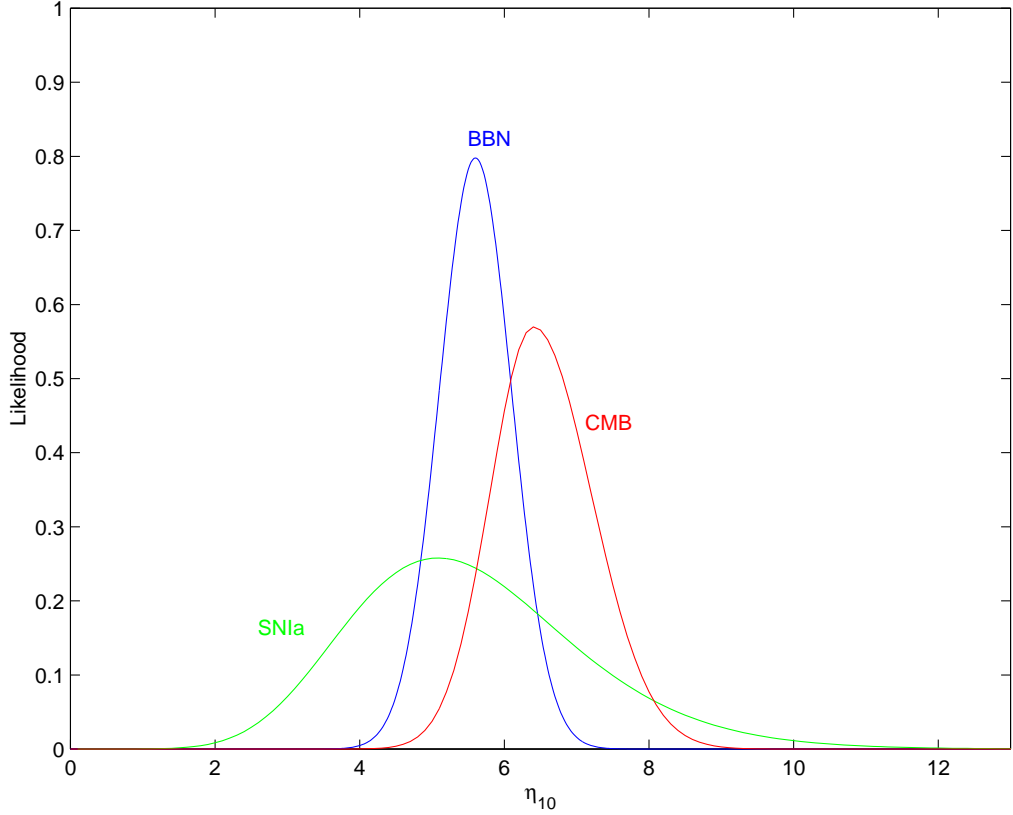}}  
\hfill  
\parbox[b]{\hsize}{
\caption{The likelihood distributions, normalized to 
equal areas under the curves, for the baryon-to-photon 
ratios ($\eta_{10}$) derived from BBN ($\sim 20$ 
minutes), from the CMB ($\sim $ few hundred thousand 
years), and for the present universe ($t_0 \sim 10$~Gyr; 
$z ~\la 1$).}}  
\end{figure}  

In Figure 12 are shown the likelihood distributions
for the three baryon density determinations discussed
above.  It is clear that these disparate determinations,
relying on completely different physics and from widely
separated epochs in the evolution of the universe are
in excellent agreement, providing strong support for
the standard, hot big bang cosmological model and for
the standard model of particle physics.  Although it 
has been emphasized many times in these lectures that
the errors are likely dominated by evolutionary and
systematic uncertainties and, therefore, are almost
certainly not normally distributed, it is hard to 
avoid the temptation to combine these three independent
estimates.  Succumbing to temptation: $\eta_{10} = 
5.8 ^{+0.4}_{-0.6}$ (\obh ~$= 0.021 ^{+0.0015}_{-0.0020}$). 

\subsection{Testing The Consistency Of SBBN}

\begin{figure}[t!]\label{hevsdobs}  
\resizebox{\hsize}{!}{\includegraphics{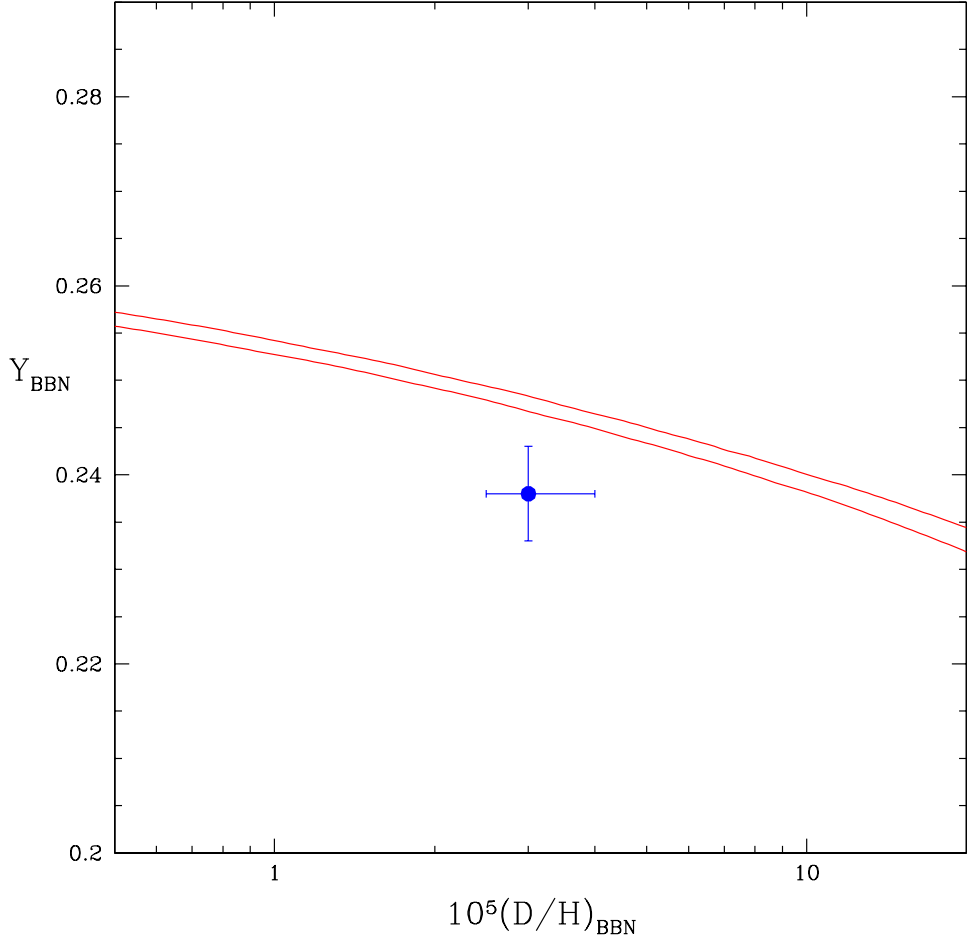}}  
\hfill  
\parbox[b]{\hsize}{
\caption{The diagonal band is the SBBN-predicted helium-4
mass fraction versus the SBBN-predicted deuterium abundance
(by number relative to hydrogen).   The width of the band 
accounts for the theoretical uncertainties in the SBBN 
predictions.  Also shown by the filled circle and error 
bars are the primordial \4he and D abundance estimates 
adopted here.}}  
\end{figure}  

As impressive as is the agreement among the three independent
estimates of the universal baryon density, we should not be 
lured into complacency.  The apparent success of SBBN should 
impell us to test the standard model even further.  How else 
to expose possible systematic errors which have heretofore 
been hidden from view or, to find the path beyond the standard 
models of cosmology and particle physics?  To this end, in 
Figure 13 are compared the SBBN \4he (Y) and D (D/H) abundance 
predictions along with the estimates for Y and D/H adopted 
here.  The agreement is not very good.  Indeed, while for 
$\eta_{10} = 5.8 ^{+0.4}_{-0.6}$, (D/H)$_{\rm SBBN} = 2.8 
^{+0.4}_{-0.7} \times 10^{-5}$, in excellent agreement with 
the O'Meara \etal (2001) estimate, the corresponding \4he 
abundance is predicted to be Y$_{\rm SBBN} = 0.248 \pm 0.001$, 
which is 2$\sigma$ above our OSW-adopted primordial abundance.  
Indeed, the SBBN-predicted \4he abundance based directly on 
deuterium is also $2\sigma$ above the IT/ITL estimate.  Here 
is a potential challenge to the internal consistency of SBBN. 
Given that systematic errors dominate, it is difficult to 
decide how seriously to take this challenge.  In fact, if 
\4he and D, in concert with SBBN, are each employed as 
baryometers, their likelihood distributions for $\eta$ 
are consistent at the 7\% level. 

\begin{figure}[t!]\label{livsdobs}  
\resizebox{\hsize}{!}{\includegraphics{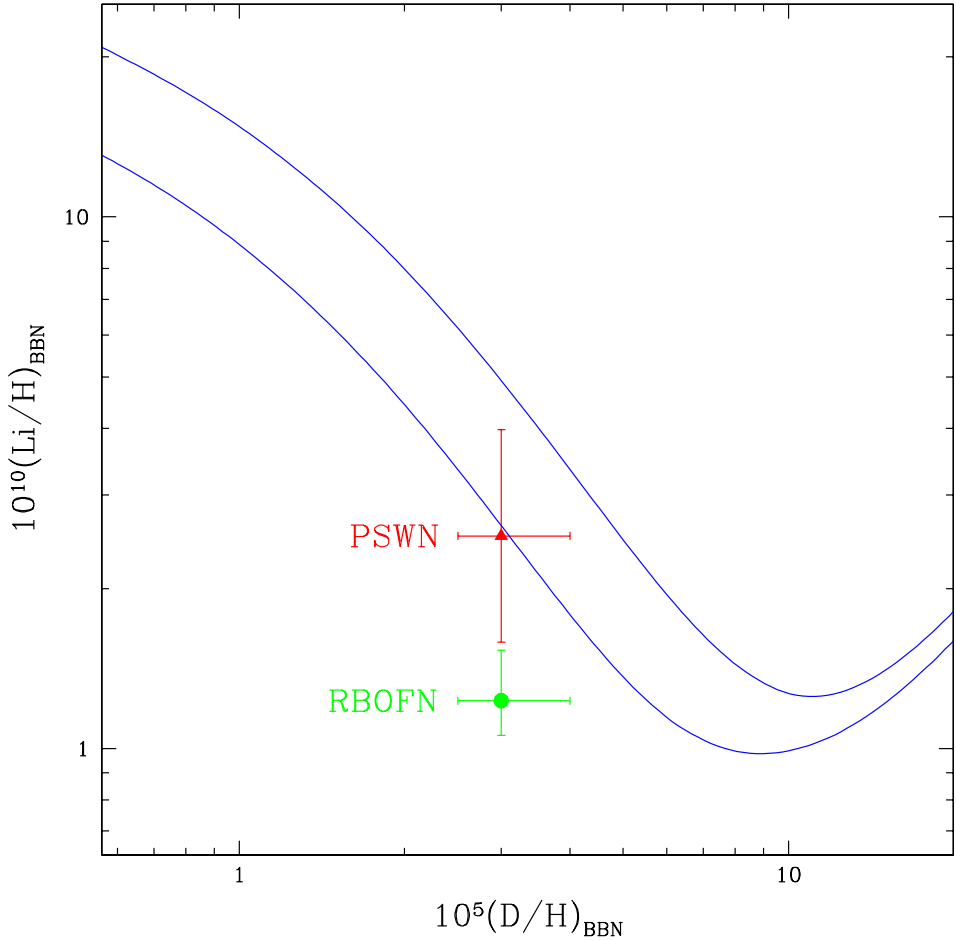}}  
\hfill  
\parbox[b]{\hsize}{
\caption{The band is the SBBN-predicted lithium abundance 
(by number relative to hydrogen) versus the SBBN-predicted
deuterium abundance.  The width of the band accounts for 
the theoretical uncertainties in the SBBN predictions.  
Also shown is the Pinsonneault \etal (2002; PSWN) primordial 
lithium abundance estimate adopted here (filled triangle) 
along with the Ryan \etal (2000; RBOFN) estimate (filled 
circle).}}  
\end{figure}  

An {\it apparent success} of (or, a {\it potential challenge} 
to) SBBN emerges from a comparison between D and \7li.  In 
Figure 14 is shown the SBBN-predicted relation between 
primordial D and primordial \7li along with the relic abundance 
estimates adopted here (Pinsonneault \etal 2002).  Also shown
for comparison is the Ryan \etal (2000) primordial lithium
abundance estimate.  The higher, depletion/dilution-corrected
lithium abundance of PSWN is in excellent agreement with the 
SBBN-D abundance, while the lower, RBOFN value poses a challenge 
to SBBN.

\section{BBN In Non-Standard Models}

As just discussed in \S4.6, there is some tension between the
SBBN-predicted abundances of D and \4he and their primordial
abundances inferred from current observational data (see Fig.~13).  
Another way to see the challenge is to superpose the data on 
the BBN predictions from Figure 2, where the \Yp versus D/H 
relations are shown for several values of N$_{\nu}$ (SSG).  
This is done in Figure 15 where it is clear that the data 
prefer {\bf nonstandard} BBN, with N$_{\nu}$ closer to 2 
than to the standard model value of 3.
  
\begin{figure}[t!]\label{hevsdNnu234obs}  
\resizebox{\hsize}{!}{\includegraphics{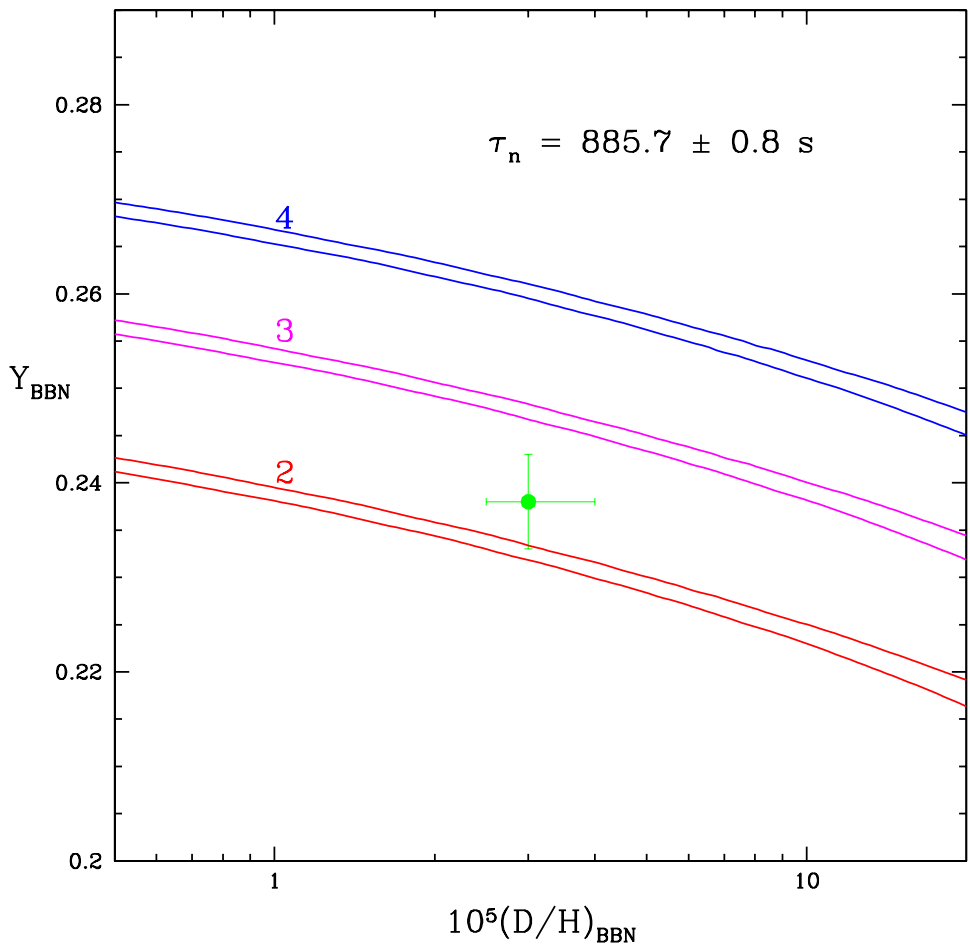}}  
\hfill  
\parbox[b]{\hsize}{  
\caption{The BBN-predicted primordial \4he mass fraction Y as 
a function of the BBN-predicted primordial Deuterium abundance  
(by number relative to Hydrogen) for three choices of N$_{\nu}$. 
Also shown by the filled circle and error bars are the primordial
abundances adopted here (\S4).}}  
\end{figure}  

It is easy to understand this result on the basis of the earlier 
discussion (see \S2.3).  The adopted abundance of D serves, 
mainly, to fix the baryon density which, in turn, determines
the SBBN-predicted \4he abundance.  The corresponding predicted 
value of \Yp is too large when compared to the data.  A universe 
which expands {\it more slowly} ($S < 1$; N$_{\nu} < 3$) will 
permit more neutrons to transmute into protons before BBN commences, 
resulting in a smaller \4he mass fraction.  However, there are 
two problems (at least!) with this ``solution''.  The main issue 
is that there {\bf are} three ``flavors'' of light neutrinos, 
so that N$_{\nu} \ge 3$ ($\Delta$N$_{\nu} \ge 0$).  The second, 
probably less serious problem is that a slower expansion permits 
an increase in the $^7$Be production, resulting in an increase 
in the predicted relic abundance of lithium.  For (D/H)$_{\rm P} 
= 3.0 \times 10^{-5}$ and \Yp = 0.238, the best fit values of 
$\eta$ and \Deln are: $\eta_{10} = 5.3$ (\obh = 0.019) and N$_{\nu} 
= 2.3$ (\Deln $ = -0.7$).  For this combination the BBN-predicted
lithium abundance is [Li]$_{\rm P} = 2.53$ ((Li/H)$_{\rm P} = 3.4 
\times 10^{-10}$), somewhat higher than, but still in agreement 
with the PSWN estimate of [Li]$_{\rm P} = 2.4 \pm 0.2$, but much 
higher than the RBOFN value of [Li]$_{\rm P} = 2.1 \pm 0.1$.  
Although the tension between the observed and SBBN-predicted 
lithium abundances may not represent a serious challenge (at 
present), the suggestion that \Deln $< 0$ must be addressed.  
One possibility is that the slower expansion of the 
radiation-domianted early universe could result from a 
non-minimally coupled scalar field (``extended quintessence'') 
whose effect is to change the effective gravitational constant 
($G \rightarrow G' < G$; see \S2.3).  For a discussion of such 
models and for further references see, \eg Chen, Scherrer \& 
Steigman (2001). 

\subsection{Degenerate BBN}

\begin{figure}[t!]\label{ksswisoabund}  
\resizebox{\hsize}{!}{\includegraphics{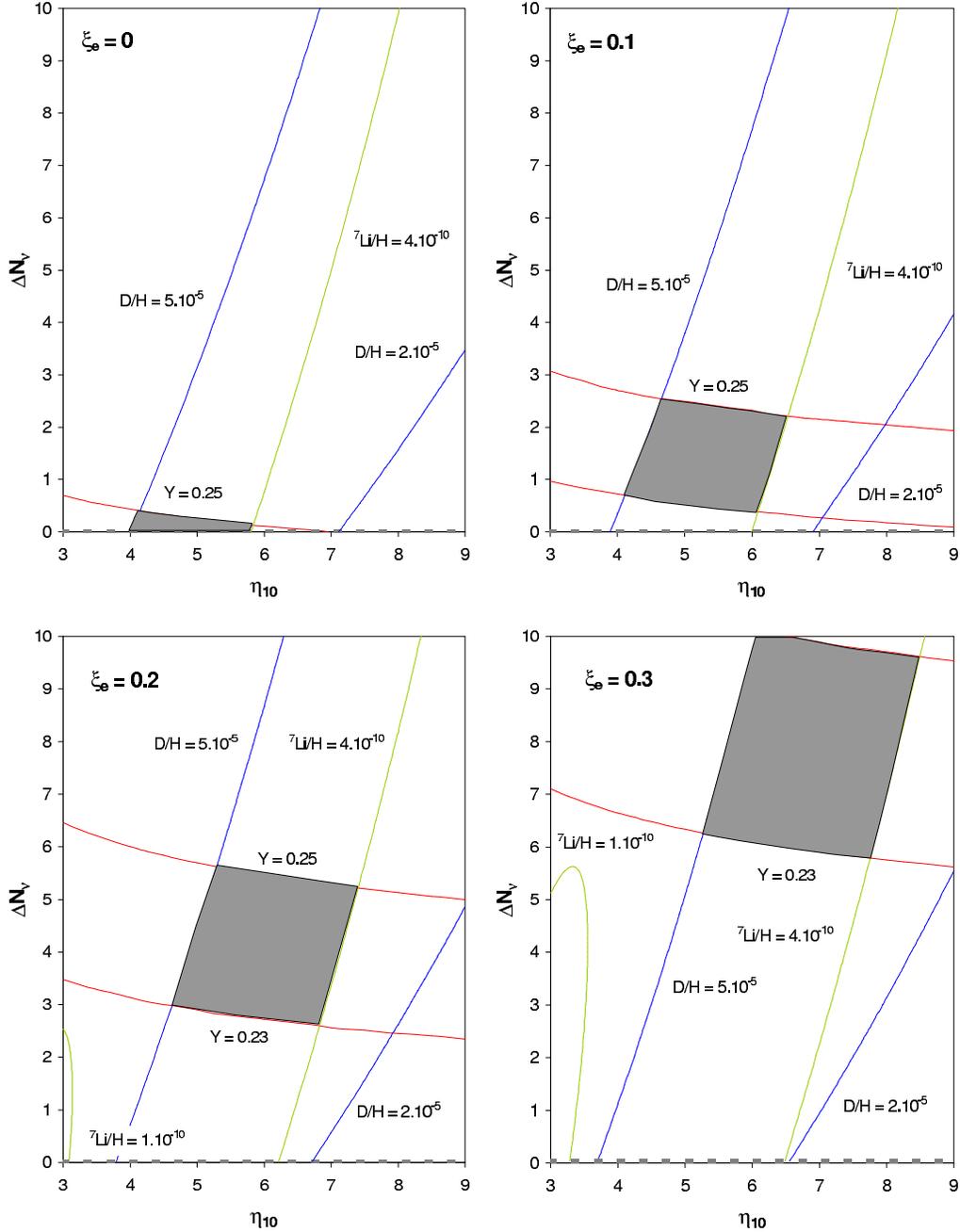}}  
\hfill  
\parbox[b]{\hsize}{  
\caption{Isoabundance contours for D, \4he, and \7li in the 
\Deln -- $\eta_{10}$ plane for four choices of the $\nu_{e}$ 
degeneracy parameter $\xi_{e}$.  The shaded areas identify
the ranges of parameters consistent with the abundances
indicated (see the text).}}  
\end{figure}  

There is another alternative to SBBN which, although currently 
less favored, does have a venerable history: BBN in the presence 
of a background of {\bf degenerate} neutrinos.  First, a brief 
diversion to provide some perspective.  In the very early
universe there were a large number of particle-antiparticle
pairs of all kinds.  As the baryon-antibaryon pairs or, 
their quark-antiquark precursors, annihilated, only the 
baryon {\it excess} survived.  This baryon number excess, 
proportional to $\eta$, is very small ($\eta ~\la 10^{-9}$).  
It is reasonable, but by no means compulsory, to assume 
that the lepton number {\it asymmetry} (between leptons and
antileptons) is also very small.  Charge neutrality of the
universe ensures that the electron asymmetry is of the same
order as the baryon asymmetry.  But, what of the asymmetry
among the several neutrino flavors?  

Since the relic neutrino background has never been observed 
directly, not much can be said about its asymmetry.  However, 
if there is an excess in the number of neutrinos compared to 
antineutrinos (or, vice-versa), ``neutrino degeneracy'', the 
total energy density in neutrinos (plus antineutrinos) is 
increased.  As a result, during the early, radiation-dominated
evolution of the universe, $\rho \rightarrow \rho' > \rho$, 
and the universal expansion rate increases ($S > 1$).  
Constraints on how large $S$ can be do lead to some weak 
bounds on neutrino degeneracy (see, \eg Kang \& Steigman 1992 
and references therein).  This effect occurs for degeneracy in
all neutrino flavors ($\nu_{e}$, $\nu_{\mu}$, and $\nu_{\tau}$).  
For {\bf fixed} baryon density, $S > 1$ leads to an increase in 
D/H (less time to destroy D), more \4he (less time to transform 
neutrons into protons), and a decrease in lithium (at high $\eta$ 
there is less time to produce $^7$Be).  Recall that for $S = 1$ 
(SBBN), an increase in $\eta$ results in less D (more rapid 
destruction), which can compensate for $S > 1$.  Similarly, 
an increase in baryon density will increase the lithium yield 
(more rapid production of $^7$Be), also tending to compensate 
for $S > 1$.  But, at higher $\eta$, more \4he is produced, 
further exacerbating the effect of a more rapidly expanding 
universe.  

However, electron-type neutrinos play a unique role in BBN, 
mediating the neutron-proton transformations via the weak 
interactions (see eq.~2.24).  Suppose, for example, there 
are more $\nu_{e}$ than $\bar{\nu}_{e}$.  If $\mu_{e}$ is
the $\nu_{e}$ chemical potential, then $\xi_{e} \equiv 
\mu_{e}/kT$ is the ``neutrino degeneracy parameter''; in
this case, $\xi_{e} > 0$.  The excess of $\nu_{e}$ will
drive down the neutron-to-proton ratio, leading to a {\it 
reduction} in the primordial \4he mass fraction.  Thus, a
combination of {\bf three} adjustable parameters, $\eta$, 
$\Delta$N$_{\nu}$, and $\xi_{e}$ may be varied to ``tune''
the primordial abundances of D, \4he, and \7li.  In Kneller
\etal (2001; KSSW), we chose a range of primordial abundances
similar to those adopted here ($2 \le 10^{5}$(D/H)$_{\rm P} 
\le 5$; $0.23 \le$ Y$_{\rm P} \le 0.25$; $1 \le 10^{10}
$(Li/H)$_{\rm P} \le 4$) and explored the consistent ranges 
of $\eta$, $\xi_{e} \ge 0$, and $\Delta$N$_{\nu} \ge 0$.  
Our results are shown in Figure 16.

It is clear from Figure 16 that for a large range in $\eta$, 
a combination of \Deln and $\xi_{e}$ {\bf can} be found so 
that the BBN-predicted abundances will lie within our adopted 
primordial abundance ranges.  However, there are constraints 
on $\eta$ and \Deln from the CMB temperature fluctuation 
spectrum (see KSSW for details and further references).  
Although the CMB temperature fluctuation spectrum is 
insensitive to $\xi_{e}$, it will be modified by any changes 
in the universal expansion rate.  While SBBN (\Deln = 0) is 
consistent with the combined constraints from BBN and the 
CMB (see \S4.5) for $\eta_{10} \approx 5.8$ (\obh $ \approx 
0.021$), values of \Deln as large as \Deln $\la 6$ are also 
allowed (KSSW).

\section{Summary}

As observations reveal, the present universe is filled with
radiation and is expanding.  According to the standard, hot
big bang cosmological model the early universe was hot and
dense and, during the first few minutes in its evolution,
was a primordial nuclear reactor, synthesizing in significant
abundances the light nuclides D, \3he, \4he, and \7li.  These
relics from the Big Bang open a window on the early evolution 
of the universe and provide probes of the standard models of 
cosmology and of particle physics. Since the BBN-predicted
abundances depend on the competition between the early universe
expansion rate and the weak- and nuclear-interaction rates,
they can be used to test the standard models as well as to
constrain the universal abundances of baryons and neutrinos.
This enterprise engages astronomers, astrophysicists, cosmologists,
and particle physicists alike.  A wealth of new observational
data has reinvigorated this subject and stimulated much recent
activity.  Much has been learned, revealing many new avenues 
to be explored (a key message for the students at this school 
-- and for young researchers everywhere).  The current high
level of activity ensures that many of the detailed, quantitative
results presented in these lectures will need to be revised 
in the light of new data, new analyses of extant data, and 
new theoretical ideas.  Nonetheless, the underlying physics
and the approaches to confronting the theoretical predictions
with the observational data presented in these lectures should 
provide a firm foundation for future progress.

Within the context of the standard models of cosmology and 
of particle physics (SBBN) the relic abundances of the light 
nuclides depend on only one free parameter, the baryon-to-photon 
ratio (or, equivalently, the present-universe baryon density 
parameter).  With one adjustable parameter and three relic
abundances (four if \3he is included), SBBN is an overdetermined
theory, potentially falsifiable.  The current status of the
comparison between predictions and observations reviewed here
illuminates the brilliant success of the standard models.
Among the relic light nuclides, deuterium is the baryometer
of choice.  For N$_{\nu} = 3$, the SBBN-predicted deuterium 
abundance agrees with the primordial-D abundance derived 
from the current observational data for $\eta_{10} = 5.6^
{+0.6}_{-1.2}$ ($\Omega_{\rm B} = 0.020^{+0.002}_{-0.004}$).  
This baryon abundance, from the first 20 minutes of the 
evolution of the universe, is in excellent agreement with
independent determinations from the CMB ($\sim $~few hundred
thousand years) and in the present universe ($\sim $~10 Gyr).

It is premature, however, to draw the conclusion that the
present status of the comparison between theory and data 
closes the door on further interesting theoretical and/or 
observational work.  As discussed in these lectures, there 
is some tension between the SBBN-predicted abundances and 
the relic abundances derived from the observational data.  
For the deuterium-inferred SBBN baryon density, the expected 
relic abundances of \4he and \7li are somewhat higher than 
those derived from current data.  The ``problems'' may lie 
with the data (large enough data sets? underestimated errors?) 
or, with the path from the data to the relic abundances 
(systematic errors? evolutionary corrections?).  For example,
has an overlooked correction to the \hii region-derived \4he
abundances resulted in a value of \Yp which is systematically
too small (\eg underlying stellar absorption)?  Are there
systematic errors in the absolute level of the lithium
abundance on the Spite Plateau or, has the correction for
depletion/dilution been underestimated?   In these lectures
the possibility that the fault may lie with the cosmology 
was also explored.  In one simple extension of SBBN, the early
universe expansion rate is allowed to differ from that in
the standard model.  It was noted that to reconcile D, and 
\4he would require a {\it slower} than standard expansion
rate, difficult to reconcile with simple particle physics
extensions beyond the standard model.  Furthermore, if 
this should be the resolution of the tension between D
and \4he, it would exacerbate that between the predicted
and observed lithium abundances.  The three abundances
could be reconciled in a further extension involving 
neutrino degeneracy (an asymmetry between electron neutrinos
and their antiparticles).  But, three adjustable parameters
to account for three relic abundances is far from satisfying.  
Clearly, this active and exciting area of current research 
still has some surprises in store, waiting to be discovered 
by astronomers, astrophysicists, cosmologists and particle 
physicists.  The message to the students at this school -- 
and those everywhere -- is that much interesting observational 
and theoretical work remains to be done.  I therefore 
conclude these lectures with a personal list of questions 
I would like to see addressed.
 
\begin{itemize}

\item
Where (at what value of D/H) is the primordial deuterium 
plateau, and what is(are) the reason(s) for the currently 
observed spread among the high-$z$, low-Z QSOALS D-abundances?

\item
Are there stellar observations which could offer complementary
insights to those from \hii regions on the question of the
primordial \4he abundance, perhaps revealing unidentified or
unquantified systematic errors in the latter approach?  Is
\Yp closer to 0.24 or 0.25?

\item
What is the level of the Spite Plateau lithium abundance?
Which observations can pin down the systematic corrections
due to model stellar atmospheres and temperature scales and
which may reveal evidence for, and quantify, early-Galaxy 
production as well as stellar depletion/destruction?

\item
If further observational and associated theoretical work
should confirm the current tension among the SBBN-predicted
and observed primordial abundances of D, \4he, \7li, what
physics beyond the standard models of cosmology and particle
physics has the potential to resolve the apparent conflicts?
Are those models which modify the early, radiation-dominated
universe expansion rate consistent with observations of the 
CMB temperature fluctuation spectrum?  If neutrino degeneracy 
is invoked, is it consistent with the neutrino properties 
(masses and mixing angles) inferred from laboratory experiments 
as well as the solar and cosmic ray neutrino oscallation data?

\end{itemize}

To paraphrase Spock, work long and prosper!

\begin{acknowledgments}  

It is with heartfelt sincerity that I thank the organizers 
(and their helpful, friendly, and efficient staff) for all 
their assistance and hospitality.  Their thoughtful planning 
and cheerful attention to detail ensured the success of this 
school. It is with much fondness that I recall the many fruitful 
interactions with the students and with my fellow lecturers 
and I thank them too.  I would be remiss should I fail to
acknowledge the collaborations with my OSU colleagues, 
students, and postdocs whose work has contributed to the 
material presented in these lectures.  The DOE is gratefully 
acknowledged for support under grant DE-FG02-91ER-40690. 
\end{acknowledgments}

\end{document}